\documentclass[12pt,onecolumn,oneside,a4paper,peerreview]{IEEEtran_kit}
\usepackage{cite}
\usepackage{float}
\usepackage{graphicx}
\usepackage{amsmath}
\usepackage{times}
\usepackage{graphicx}
\usepackage{latexsym}
\usepackage{graphicx}
\usepackage{bm}
\usepackage{amssymb}
\usepackage[center]{caption2}
\usepackage{stfloats}
\usepackage{cases}
\usepackage{array}
\usepackage{setspace}
\usepackage{fancyhdr}
\usepackage{color}
\usepackage{subfigure}
\usepackage{citesort}
\usepackage{multirow}
\usepackage{amsmath}
\usepackage{cases}
\usepackage[noend]{algpseudocode}
\usepackage{algorithm}
\usepackage{algorithmicx}
\usepackage{diagbox}
\usepackage{algpseudocode}
\usepackage{epstopdf}
\usepackage{varwidth}
\usepackage{pifont}

\usepackage{verbatim}

\usepackage[colorlinks,urlcolor=blue,linkcolor=black]{hyperref}

\usepackage{url}

\def\bm#1{\mbox{\boldmath $#1$}}

\renewcommand{\baselinestretch}{0.97}

\begin{document}

\title{Deep Learning-based Channel Estimation for Wideband Hybrid MmWave Massive MIMO}

 \author{\authorblockN{Jiabao Gao,~\IEEEmembership{Graduate Student Member,~IEEE,} Caijun Zhong,~\IEEEmembership{Senior Member,~IEEE,} Geoffrey Ye Li,~\IEEEmembership{Fellow,~IEEE,} Joseph B. Soriaga, and Arash Behboodi,~\IEEEmembership{Member,~IEEE}
 \thanks{J. Gao and C. Zhong are with the College of Information Science and Electronic Engineering, Zhejiang University, Hangzhou 310007, China (Email: \{gao\_jiabao, caijunzhong\}@zju.edu.cn).}
 \thanks{G. Y. Li is with the Department of Electrical and Electronic Engineering, Imperial College London, London SW7 2BU, U.K. (Email: Geoffrey.Li@imperial.ac.uk).}
 \thanks{J. B. Soriaga is with Qualcomm Technologies, Inc., 5775 Morehouse Dr, San Diego CA 92122. (Email: jsoriaga@qti.qualcomm.com).}
\thanks{A. Behboodi is with Qualcomm Technologies Netherlands B.V., Science Park 400, 1098 XH Amsterdam. (Email: behboodi@qti.qualcomm.com).}
 }}

\maketitle

\begin{abstract}
Hybrid analog-digital (HAD) architecture is widely adopted in practical millimeter wave (mmWave) massive multiple-input multiple-output (MIMO) systems to reduce hardware cost and energy consumption. However, channel estimation in the context of HAD is challenging due to only limited radio frequency (RF) chains at transceivers. Although various compressive sensing (CS) algorithms have been developed to solve this problem by exploiting inherent channel sparsity and sparsity structures, practical effects, such as power leakage and beam squint, can still make the real channel features deviate from the assumed models and result in performance degradation. Also, the high complexity of CS algorithms caused by a large number of iterations hinders their applications in practice. To tackle these issues, we develop a deep learning (DL)-based channel estimation approach where the sparse Bayesian learning (SBL) algorithm is unfolded into a deep neural network (DNN). In each SBL layer, Gaussian variance parameters of the sparse angular domain channel are updated by a tailored DNN, which is able to effectively capture complicated channel sparsity structures in various domains. Besides, the measurement matrix is jointly optimized for performance improvement. Then, the proposed approach is extended to the multi-block case where channel correlation in time is further exploited to adaptively predict the measurement matrix and facilitate the update of Gaussian variance parameters. Based on simulation results\footnote{The source code is available at \href{https://github.com/EricGJB/SBL_unfolding_based_HAD_MIMO_Channel_Estimation}{$\mathrm{https://github.com/EricGJB/SBL\_unfolding\_based\_HAD\_MIMO\_Channel\_Estimation}$}.}, the proposed approaches significantly outperform existing approaches but with reduced complexity.

\newpage
\begin{center}
{\bf Index Terms}
\end{center}
Millimeter wave, massive MIMO, hybrid analog-digital, wideband channel estimation, measurement matrix, power leakage, beam squint, sparse Bayesian learning, deep learning, deep unfolding.
\end{abstract}


\section{Introduction}
Future wireless communication systems require extremely high transmission rate to realize various newly emerging applications, such as high quality three-dimensional (3D) video, virtual reality, and augmented reality\cite{towards_6G}. By deploying a large number of antennas at transceivers, massive multiple-input multiple-output (MIMO) is able to dramatically improve the spectral efficiency thanks to the extra degrees of freedom in the spatial domain\cite{massive_mimo2}. On the other hand, millimeter wave (mmWave) communication is envisioned as another key enabling technology for higher throughput with huge available bandwidth\cite{mmWave2}. Thanks to the naturally complementary features of these two technologies, mmWave band is ideally suited for the deployment of massive MIMO. Specifically, the short wavelength of mmWave enables the integration of massive antenna units in a limited space while its severe path loss and signal blockage can be compensated by the high beamforming gain of massive MIMO. Nevertheless, mmWave massive MIMO faces the practical challenge of prohibitive manufacturing cost and energy consumption of a large number of radio frequency (RF) chains. To alleviate this issue, the hybrid analog-digital (HAD) architecture has been proposed where antennas are connected to much fewer RF chains through phase shifters in the analog domain and the performance is further enhanced by digital processing\cite{had2}.

It is well known that various gains of massive MIMO heavily rely on accurate channel state information (CSI). In fully-digital systems, linear estimators like least square (LS) can be used to acquire the CSI. However, channel estimation is much harder with HAD since the high-dimensional channel needs to be recovered from the low-dimensional compressed pilot signal. To achieve the same performance as in fully-digital systems, the estimation overhead of linear estimators will increase dramatically since only part of the channel can be estimated at once with limited RF chains. Therefore, the development of channel estimation algorithms with both high accuracy and low overhead is critical to advance the deployment of HAD massive MIMO. Next, we will present a comprehensive review of prior works in this context.

\subsection{Review of Prior Works}
The key idea of the most existing methods is to leverage the inherent sparsity and sparsity structures of the channel. Due to limited scattering caused by the highly directional propagation behavior of mmWave, only few channel paths can reach the receiver. Therefore, the number of effective channel parameters in the angular domain is much smaller than that of antenna pairs of transceivers\cite{svbi}. To reduce the overhead of LS, in \cite{reduced_LS}, the complete channel is first estimated in the preamble stage to obtain slowly changing path angles, then only path gains need to be re-estimated for a long period. Nevertheless, the initial preamble can be resource-consuming with limited RF chains and this method is not robust to sudden change of path angles. 

The compressive sensing (CS) technique is commonly used to exploit the channel sparsity in prior works, where channel estimation is formulated as a sparse recovery problem and effective channel parameters are estimated instead of the original channel\cite{omp_ce,somp_ce,sbl_ce,pc_sbl_2d_ce,s_sbl,amp_3d,HAD_MIMO,off_grid_sbl_ce,ESPRIT_ce,beam_squint_ce}. In \cite{omp_ce}, the orthogonal matching pursuit (OMP) algorithm is used to estimate the sparse angular domain channel. Later on, an improved version is proposed in \cite{somp_ce} by further exploiting the common sparsity structure among channels corresponding to different subcarriers. In \cite{sbl_ce}, the sparse Bayesian learning (SBL) algorithm is proposed for channel estimation and two of its variants are developed to exploit the channel correlation in the consecutive blocks. The parameterized Gaussian priors of SBL are designed to capture the block sparsity structure in the clustered mmWave channel\cite{pc_sbl_2d_ce} and the one-ring low-frequency channel\cite{s_sbl}. Similar idea is adopted in \cite{amp_3d}, where approximate message passing (AMP) with nearest neighbor pattern learning is proposed to exploit the 3D clustered sparsity structure in the AoA-AoD-delay domain. In \cite{svbi}, variational Bayesian inference (VBI) algorithms are proposed to improve estimation performance at the cost of high complexity. 

In spite of reduced overhead, the performance of CS algorithms heavily relies on the sparsity assumptions and will be severely degraded when real channel features deviate from the assumed models. For instance, the power leakage effect caused by grid mismatch will decrease the sparsity level of channel\cite{reduced_LS} while the beam squint effect in wideband mmWave massive MIMO systems will damage the common sparsity among different subchannels\cite{beam_squint_ce}. To mitigate these effects, grid refinement is performed on initial estimated grids with a higher resolution dictionary in \cite{HAD_MIMO}. In \cite{off_grid_sbl_ce}, an off-grid SBL algorithm is proposed to jointly estimate grid mismatch parameters with channel parameters. Two ESPRIT-based algorithms \cite{ESPRIT_ce} are developed for high-resolution channel estimation with closed-form solutions for path angles. In \cite{beam_squint_ce}, the majorization-minimization iterative approach is used to estimate squint-independent time domain channel parameters. On the other hand, CS algorithms are also often criticized for their high complexity since a large number of iterations are required for convergence.

Deep learning (DL) has recently attracted widespread attention in wireless communication due to its strong representation ability and low testing complexity. Up to now, DL has been successfully applied to many physical layer problems\cite{DL_physical}, such as CSI feedback\cite{DL_CSI_feedback}, resource allocation\cite{DL_RA}, beamforming\cite{DL_BF}, signal detection\cite{DL_SD1}, end-to-end transceiver design\cite{DL_E2E}, and channel estimation\cite{cnn_refine,fdd_autoencoder,attention,DLCS,somp_refine,LDAMP,beam_squint_hht,beam_squint_LISTA,mmv_amp,off_grid_sbl_unfold}. DL-based channel estimation methods can be generally divided into two categories, namely data-driven ones and model-driven ones. 

Data-driven methods directly approximate the mapping function from received signal or estimation to channel or effective channel parameters by a black-box deep neural network (DNN). In \cite{cnn_refine}, a convolutional neural network (CNN) has been proposed to refine the coarse LS estimation in HAD massive MIMO systems where channel correlations in the frequency and time domains are exploited for performance improvement and overhead reduction. The initial coarse estimation can also be obtained by a conventional CS algorithm\cite{somp_refine} or a fully-connected network (FNN)\cite{fdd_autoencoder}. In \cite{attention}, the attention mechanism is embedded into the FNN to realize divide-and-conquer by exploiting the separability of channel distribution, which achieves better performance than VBI but is with lower complexity. In \cite{DLCS}, an FNN is trained to predict the amplitudes of beamspace channel entries, which is extended to the multi-carrier case in \cite{HAD_MIMO} using CNN. 

Different from data-driven methods, model-driven methods combine the domain knowledge in communication with the power of DL by inserting trainable parameters into the conventional algorithms, therefore usually having better interpretability and robustness. In \cite{LDAMP}, the AMP algorithm is unfolded with a deep CNN-based channel denoiser for beamspace mmWave massive MIMO channel estimation, which is extended to the system with beam squint in \cite{beam_squint_hht}. To fully exploit channel sparsity, a learnable iterative shrinkage thresholding algorithm (LISTA) is proposed in \cite{beam_squint_LISTA} where channel denoising is further performed in the sparse transform domain. In \cite{mmv_amp}, deep unfolding is conducted on the multiple-measurement-vectors (MMV) version of AMP where the common sparsity among subchannels is embedded. In \cite{off_grid_sbl_unfold}, learnable parameters are inserted into the off-grid SBL algorithm to accelerate convergence without harming performance.

Another important part to design in sparse recovery problems is the measurement matrix. Conventional matrices are usually designed to satisfy restricted isometry property with high probability, have low mutual coherence, or achieve almost equal received signal power for all angles\cite{s_sbl,svbi,ESPRIT_ce}. In \cite{mmv_amp,fdd_autoencoder,location_aided} the matrix elements are treated as trainable network parameters and the optimal matrix in terms of average estimation error on a given dataset can be obtained.

\subsection{Motivations and Contributions}
Most of current DL-based channel estimation works have not fully exploited channel sparsity structures, especially when the structures are hard to be expressed mathematically. Besides, the jointly optimized measurement matrix is long-term fixed and not flexible enough to match the time-varying short-term channel distribution. In \cite{l_sbl}, an SBL unfolding framework has been proposed to learn arbitrary sparsity structures. In spite of the inspiring idea, the FNN used therein performs badly in our channel estimation problem since the local correlation in the structured channel can not be effectively captured, and the complexity could be very high in large scale systems. In \cite{location_aided}, adaptive measurement matrices are learned by training multiple DNNs with the aid of location information. However, this method relies on the separability of channel distribution and is only applicable to low-frequency channel with narrow angular spread.

To overcome the shortcomings of existing approaches, in this paper, we propose a DL-based channel estimation approach for wideband HAD mmWave massive MIMO systems. The main contributions of this paper are as follows:
\begin{itemize}
\item
An SBL unfolding-based channel estimation approach is proposed, where the SBL algorithm with multiple iterations is unfolded into a DNN with cascaded SBL layers, and the update of Gaussian variance parameters in each SBL layer is realized by a tailored DNN\footnote{To avoid ambiguity, in this sentence, the first DNN refers to the overall network containing all SBL layers while the second DNN refers to the dedicated network for variance parameter update in each SBL layer.}. Besides, the measurement matrix is jointly optimized to improve measurement efficiency.
\item
The sparsity structures of wideband mmWave channel with practical effects including power leakage and beam squint are analyzed. Then, the architecture of the DNN for variance parameter update in each SBL layer is carefully designed according to channel sparsity structures in various domains to achieve both excellent performance and low complexity.
\item 
The proposed approach is extended to the multi-block case by further exploiting channel correlation in the time domain. Based on previous estimation results, the measurement matrix matching the time-varying short-term channel sparsity patterns is adaptively predicted and the update of variance parameters is facilitated.
\end{itemize}

\subsection{Organization and Notations}
The rest of this paper is organized as follows. In Section II, the HAD massive MIMO system, the wideband mmWave channel model, and the uplink channel estimation problem are introduced. In Section III, conventional SBL-based channel estimation approaches are presented. Section IV elaborates the proposed DL-based channel estimation approach, which is further extended to the multi-block case in Section V. Simulation results are provided in Section VI to validate the superiority of the proposed approaches, and the paper is concluded in Section VII.

\emph{Notations:} Italic, bold-face lower-case and bold-face upper-case letters denote scalar, vector, and matrix, respectively. ${\left\| {\cdot} \right\|}$, $(\cdot)^{*}$, $(\cdot)^{T}$, $(\cdot)^{H}$, $(\cdot)^{-1}$, $|\cdot|$, $\mathbb{E}\{\cdot\}$, and $\otimes$ denote $l$-2 norm, conjugate, transpose, conjugate transpose, inverse, modulus, expectation, and Kronecker product, respectively. $\bm{I}_a$ denotes the $(a,a)$-dimensional identity matrix. $[\cdot]_{i,j}$, $[\cdot]_{i.}$, and $[\cdot]_{.j}$ denote the element at the $i$-th row and $j$-th column, the elements at the $i$-th row, and the elements at the $j$-th column of a matrix, respectively. $\mathrm{vec}(\cdot)$ denotes the vectorization of a matrix by stacking its columns while $\mathrm{vec}^{-1}(\cdot)$ denotes the inverse operation. $\mathrm{diag(\cdot)}$ denotes both the diagonal matrix and diagonalization operation. $\mathrm{Blkdiag}(\cdot)$ denotes the block diagonal matrix. ${\mathbb C^{x \times y}}$ and ${\mathbb R^{x \times y}}$ denote the ${x \times y}$ complex and real spaces, respectively. $\mathcal{CN}(\mu,\sigma^2)$ denotes a circularly symmetric complex Gaussian (CSCG) random variable with mean $\mu$ and variance $\sigma^2$ while $\mathcal{CN}(\bm{\mu},\bm{\Sigma})$ denotes a CSCG random vector with mean $\bm{\mu}$ and covariance $\bm{\Sigma}$. $\mathcal{U}[a,b]$ denotes the uniform distribution between $a$ and $b$.

\section{System model and problem formulation}
In this section, we first introduce the system model and channel model. Then, the sparsity structures of channel are analyzed. At last, the channel estimation problem is formulated. 

\subsection{System Model}
\begin{figure}[!htb]
\centering
\includegraphics[width=1\textwidth]{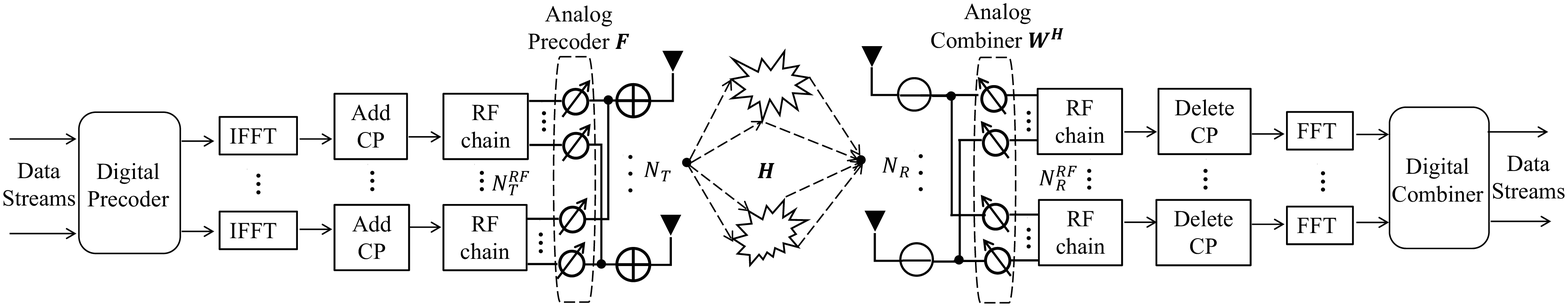}
\caption{HAD massive MIMO system.}
\label{system}
\end{figure}

Consider the HAD massive MIMO system illustrated in Fig. \ref{system}, where a base station (BS) with an $N_R$-antenna uniform linear array (ULA) and $N_R^{RF}$ RF chains serves a user with an $N_T$-antenna ULA and $N_T^{RF}$ RF chains. Time division duplex (TDD) is adopted for uplink and downlink transmissions so that only the uplink channel needs to be estimated and the downlink channel can be obtained based on channel reciprocity\cite{svbi}. 

Following the common practice, in each channel use of uplink channel estimation, the user only activates a single RF chain to transmit the pilot signal on one beam while the BS combines the received pilot signal using all RF chains associated with different beams\cite{cnn_refine}. Denote the number of total transmit beams and receive beams as $M_T$ and $M_R$, respectively, and assume $M_R$ is an integer multiple of $N_R^{RF}$. With a fixed transmit beam, the BS needs $\frac{M_R}{N_R^{RF}}$ channel uses to traverse all receive beams, where a group of $N_R^{RF}$ receive beams are adopted in each channel use. Then, the user changes the transmit beam every $\frac{M_R}{N_R^{RF}}$ channel uses for $M_T$ times. 

To combat the frequency selectivity of channel, the wideband OFDM system is considered, where the beams implemented by phase shifters are shared by all subcarriers and a channel use is equivalent to an OFDM symbol. In each block, the first $\frac{M_TM_R}{N_R^{RF}}$ symbols are pilots for channel estimation while the rest symbols are for data transmission. In the frequency domain, $K$ subcarriers are uniformly selected to carry pilots and the channels of the rest subcarriers carrying data symbols can be obtained by methods like interpolation.

Denote the transmit beamforming matrix and the receive combining matrix as $\bm{F}=[\bm{f}_1,\bm{f}_2,\cdots,\\ \bm{f}_{M_T}]\in \mathbb{C}^{N_T\times M_T}$ and $\bm{W}=[\bm{w}_1,\bm{w}_2,\cdots,\bm{w}_{M_R}]\in \mathbb{C}^{N_R\times M_R}$, respectively. When the user adopts the $p$-th transmit beam $\bm{f}_p\in \mathbb{C}^{N_T\times 1}$ and the BS adopts the $q$-th group of receive beams $\bm{W}_{q}=[\bm{w}_{(q-1)N_R^{RF}+1},\bm{w}_{(q-1)N_R^{RF}+2},\cdots,\bm{w}_{qN_R^{RF}}]\in \mathbb{C}^{N_R\times N_R^{RF}}$, the received signal at the BS of the $k$-th subcarrier can be expressed as\footnote{Digital beamforming and combining are not considered during channel estimation here as in \cite{cnn_refine,DLCS}.}
\begin{equation}
\bm{y}_{p,q}^k=\bm{W}_q^H(\bm{H}^k\bm{f}_ps_{p,q}^k+\bm{n}_{p,q}^k)\in \mathbb{C}^{N_R^{RF}\times 1},
\end{equation}
where $\bm{H}^k\in \mathbb{C}^{N_R\times N_T}$, $s_{p,q}^k$, and $\bm{n}_{p,q}^k\sim \mathcal{CN}(\bm{0},\sigma^2\bm{I}_{N_R})$ denote the channel matrix, the transmitted pilot signal, and the noise vector before combining of the $k$-th subcarrier, respectively. Without loss of generality, $s_{p,q}^k=1$ is assumed for all $p,q,k$ since it is known at the BS and can be readily eliminated while the signal-to-noise-ratio (SNR), defined as $1/\sigma^2$, can be adjusted by changing the noise variance $\sigma^2$. Concatenating the received signals corresponding to all receive beams with the $p$-th transmit beam, we have
\begin{equation}
\bm{y}_{p}^k=\bm{W}^H\bm{H}^k\bm{f}_p+\tilde{\bm{n}}_{p}^k\in \mathbb{C}^{M_R\times 1},
\end{equation}
where the effective noise vector and its covariance matrix are expressed as follows:
\begin{equation}
\tilde{\bm{n}}_{p}^k=[(\bm{W}_1^H\bm{n}_{p,1}^k)^T,(\bm{W}_2^H\bm{n}_{p,2}^k)^T,\cdots,(\bm{W}_{\frac{M_R}{N_R^{RF}}}^H\bm{n}_{p,\frac{M_R}{N_R^{RF}}}^k)^T]^T,
\end{equation}
\begin{equation}
\bm{R}_{\tilde{\bm{n}}_p}=\mathbb{E}[\tilde{\bm{n}}_{p}^k(\tilde{\bm{n}}_p^k)^H]=\mathrm{Blkdiag}(\sigma^2\bm{W}_1^H\bm{W}_1,\sigma^2\bm{W}_2^H\bm{W}_2,\cdots,\sigma^2\bm{W}_{\frac{M_R}{N_R^{RF}}}^H\bm{W}_{\frac{M_R}{N_R^{RF}}}).
\end{equation}

Further concatenating the received signals corresponding to all transmit beams, we have
\begin{equation}
\bm{Y}^k=\bm{W}^H\bm{H}^k\bm{F}+\tilde{\bm{N}}^k\in \mathbb{C}^{M_R\times M_T},
\label{signal_model_matrix}
\end{equation}
where $\tilde{\bm{N}}^k=[\tilde{\bm{n}}_1^k,\tilde{\bm{n}}_2^k,\cdots,\tilde{\bm{n}}_{M_T}^k]$.

\subsection{Channel Model}
In this paper, the typical clustered mmWave channel model\cite{sbl_ce,off_grid_sbl_ce} is adopted and the practical beam squint effect\cite{beam_squint_ce,wideband_beamforming} in wideband mmWave massive MIMO systems is considered. Specifically, the $k$-th uplink subchannel can be expressed as
\begin{equation}
\bm{H}^k=\sqrt{\frac{N_TN_R}{N_cN_p}}\sum_{i=1}^{N_c}\sum_{j=1}^{N_p}\alpha_{i,j}e^{-j2\pi\tau_{i}f_s\frac{k}{K}}\bm{a}_R(\theta^R_{i,j},k)\bm{a}_T(\theta^T_{i,j},k)^H,
\label{channel_model}
\end{equation}
where $N_c$ and $N_p$ denote the number of clusters and the number of subpaths in a cluster, respectively, and $\alpha_{i,j}\sim \mathcal{CN}(0,1)$, $\theta_{i,j}^R$, and $\theta_{i,j}^T$ denote the path gain, the angle of arrival (AoA) at the BS, and the angle of departure (AoD) at the user of the $j$-th subpath in the $i$-th cluster, respectively. $\tau_{i}\sim \mathcal{U}[0,\tau_{max}]$ is the delay of the $i$-th cluster with $\tau_{max}$ denoting the maximum delay. Consider half-wavelength antenna spacing, the frequency-dependent response vector of ULA\footnote{With large bandwidth, the propagation delay across the large antenna array becomes comparable to the time-domain sample period, a.k.a. the spatial-wideband effect. In frequency domain, it turns into the beam squint effect such that different subcarriers will ``see" different angles of a same channel path. Please refer to \cite{beam_squint_ce,wideband_beamforming} for detailed derivation.} at the BS side can be expressed as
\begin{equation}
\bm{a}_R(\theta_{i,j}^R,k)=\frac{1}{\sqrt{N_R}}[1,e^{-j\pi\varphi_{i,j,k}^R},\cdots,e^{-j\pi(N_R-1)\varphi_{i,j,k}^R}]^T,
\label{steeringR}
\end{equation}
where $\varphi_{i,j,k}^R\triangleq(1+\frac{kf_s}{Kf_c})\mathrm{sin}(\theta_{i,j}^R)$, $f_s$ denotes the total system bandwidth, and $f_c$ denotes the carrier frequency. Similarly, the response vector at the user side can be expressed as
\begin{equation}
\bm{a}_T(\theta_{i,j}^T,k)=\frac{1}{\sqrt{N_T}}[1,e^{-j\pi\varphi_{i,j,k}^T},\cdots,e^{-j\pi(N_T-1)\varphi_{i,j,k}^T}]^T,
\label{steeringT}
\end{equation}
where $\varphi_{i,j,k}^T\triangleq(1+\frac{kf_s}{Kf_c})\mathrm{sin}(\theta_{i,j}^T)$. Denote the mean AoA, mean AoD, angular spread of AoA, and angular spread of AoD of the $i$-th cluster as $\bar{\theta}^R_i$, $\bar{\theta}^T_i$, $\triangle\theta^R_i$, and $\triangle\theta^T_i$, respectively, we have $\theta^R_{i,j}\sim \mathcal{U}(\bar{\theta}^R_i-\triangle\theta^R_i,\bar{\theta}^R_i+\triangle\theta^R_i),\theta^T_{i,j}\sim \mathcal{U}(\bar{\theta}^T_i-\triangle\theta^T_i,\bar{\theta}^T_i+\triangle\theta^T_i),\forall i,j$, where $\bar{\theta}^R_i,\bar{\theta}^T_i\sim \mathcal{U}[0,2\pi]$ and $\triangle\theta^R_i,\triangle\theta^T_i\ll \pi$.

To facilitate channel estimation, the original channel needs to be converted to a sparse domain first, which is apparently the angular domain in this case due to limited channel clusters and narrow angular spread within clusters. Sample $G$ discrete angular grids, $\phi_1,\phi_2,\cdots,\phi_G$, from the angle space such that $\mathrm{sin}(\phi_i)=\frac{2i-1-G}{G}, \forall 1\le i\le G$, we can obtain the transmit and receive dictionary matrices as $\bm{A}_T(\phi)\triangleq [\bm{a}_T(\phi_1),\bm{a}_T(\phi_2),\cdots,\bm{a}_T(\phi_G)]\in \mathbb{C}^{N_T\times G}$ and $\bm{A}_R(\phi)\triangleq [\bm{a}_R(\phi_1),\bm{a}_R(\phi_2),\cdots,\bm{a}_R(\phi_G)]\in \mathbb{C}^{N_R\times G}$, respectively, where $\bm{a}_T(\cdot)$ and $\bm{a}_R(\cdot)$ are normal ULA response vectors with $\varphi_{i,j,k}$ reduced to $\mathrm{sin}(\theta_{i,j})$ in Equations (\ref{steeringR}) and (\ref{steeringT}). Denote the equivalent angular domain channel matrix of the $k$-th subcarrier as $\bm{X}^k\in \mathbb{C}^{G\times G}$. Since $\bm{A}_T(\phi)\bm{A}_T^H(\phi)=\frac{G}{N_T}\bm{I}_{N_T}$ and $\bm{A}_R(\phi)\bm{A}_R^H(\phi)=\frac{G}{N_R}\bm{I}_{N_R}$, the mutual conversions between $\bm{X}^k$ and $\bm{H}^k$ are as follows:
\begin{equation}
\bm{X}^k=\frac{N_TN_R}{G^2}\bm{A}_R^H(\phi)\bm{H}^k\bm{A}_T(\phi),
\end{equation}
\begin{equation}
\bm{H}^k=\bm{A}_R(\phi)\bm{X}^k\bm{A}_T^H(\phi).
\label{recover}
\end{equation}

\subsection{Channel Sparsity Structures}
\label{sparsity_structures}
\begin{figure}[htb!] 
\centering 
\vspace{-0.35cm} 
\subfigtopskip=2pt 
\subfigbottomskip=1pt 
\subfigcapskip=-3pt 
\subfigure[ $|\bm{X}^1|$]{
\label{channel1}
\includegraphics[width=0.48\textwidth]{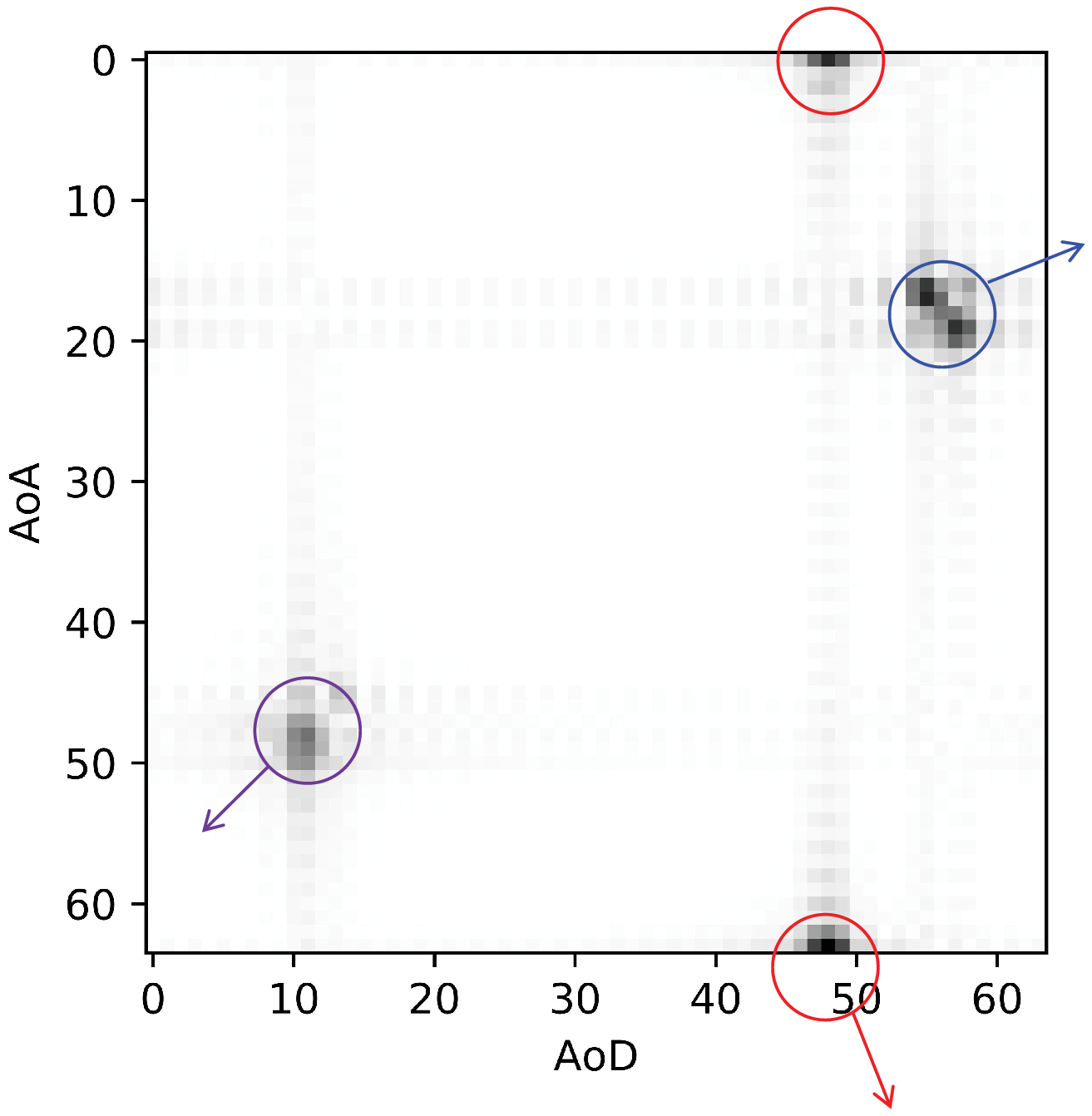}}
\subfigure[$|\bm{X}^K|$]{
\label{channel2}
\includegraphics[width=0.48\textwidth]{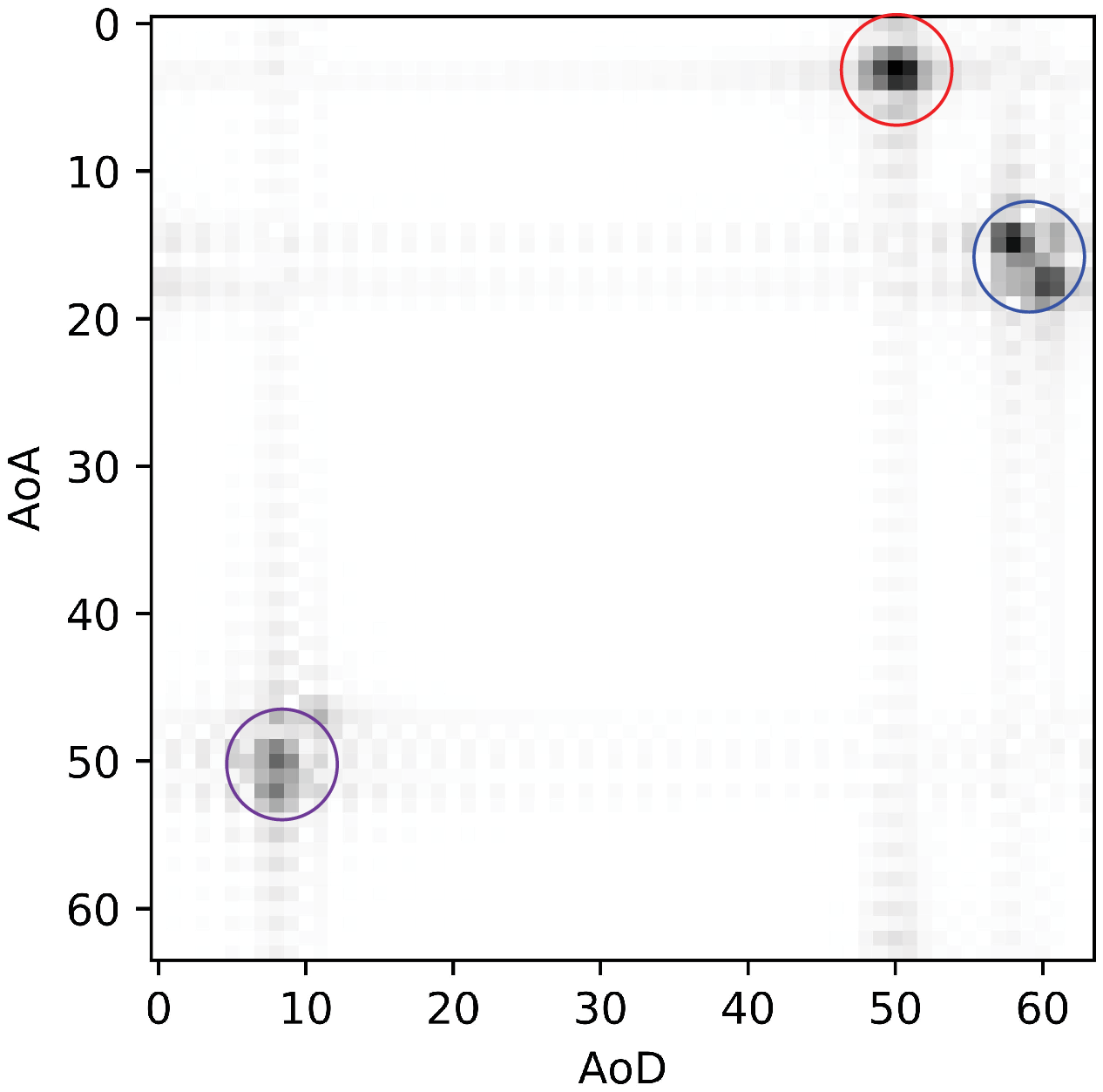}}
\caption{An example of the first and the last angular domain subchannels' modulus. In these two grayscale images, the deeper the color, the larger the modulus. The system and channel parameters here are the same as the typical setting in Section \ref{simulation}.}
\label{channel}
\end{figure}

Proper exploitation of channel sparsity structures is critical to improve channel estimation performance\cite{sbl_ce,attention}. An example of the wideband mmWave massive MIMO channel is given in Fig. \ref{channel}, from which two main sparsity structures are observed:
\begin{itemize}
\item {\bf Circular Block Sparsity Structure:} Due to the clustered distribution of channel paths, the angular domain channel exhibits block sparsity such that nearby elements tend to have close levels of modulus, as shown in the shaded areas circled in Fig. \ref{channel}. Then, the power leakage effect\footnote{The real channel angles do not necessarily lie on the predefined angular grids, and a channel path whose angle is not equal to any of the grids will have significantly non-zero responses on multiple nearby grids.} further makes the block sparsity circular, i.e., elements on the top and bottom (or the left and right) edges also have close levels of modulus. For instance, the actual AoAs of subpaths in one of the clusters of the given exemplary channel are around $90^\circ$, which correspond to the red circle on the bottom edge in Fig. \ref{channel1}, while elements in another red circle on the top edge are also significantly non-zero.
\item {\bf Imperfect Common Sparsity Structure:} The common sparsity structure among different subchannels has been widely exploited in the existing works\cite{somp_ce,mmv_amp}, which, however, becomes imperfect with the beam squint effect. Comparing Fig. \ref{channel1} and Fig. \ref{channel2}, we can see that the locations of circles corresponding to the same cluster shift between different subchannels due to frequency-dependent ULA response vectors. Furthermore, we notice that clusters near the edge in one subchannel may jump to the opposite edge in another subchannel due to the periodicity of sinusoidal function, e.g., the red circles in Fig. \ref{channel1}.
\end{itemize}

\subsection{Problem Formulation}
Stacking the columns of $\bm{Y}^k$, $\bm{H}^k$, $\bm{X}^k$, and $\tilde{\bm{N}}^k$ yield $\bm{y}^k=\mathrm{vec}(\bm{Y}^k)\in \mathbb{C}^{M_RM_T\times 1}$, $\bm{h}^k=\mathrm{vec}(\bm{H}^k)\in \mathbb{C}^{N_RN_T\times 1}$, $\bm{x}^k=\mathrm{vec}(\bm{X}^k)\in \mathbb{C}^{G^2\times 1}$, and $\bm{\tilde{\bm{n}}}^k=\mathrm{vec}(\tilde{\bm{N}}^k)\in \mathbb{C}^{M_RM_T\times 1}$, respectively. Then, exploiting the property of Kronecker product that $\mathrm{vec}(\bm{ABC})=(\bm{C}^T\otimes \bm{A})\mathrm{vec}(\bm{B}))$, we have $\bm{h}^k=(\bm{A}_T^*(\phi)\otimes \bm{A}_R(\phi))\bm{x}^k$ and $\bm{y}^k=(\bm{F}^T\otimes \bm{W}^H)\bm{h}^k+\tilde{\bm{n}}^k$. Therefore, Equation (\ref{signal_model_matrix}) can be written in the standard CS form as
\begin{equation}
\bm{y}^k=\bm{\Phi}\bm{x}^k+\tilde{\bm{n}}^k,
\label{signal_model_vector}
\end{equation}
\begin{equation}
\bm{\Phi}\triangleq (\bm{F}^T\otimes \bm{W}^H)(\bm{A}_T^*(\phi)\otimes \bm{A}_R(\phi))\in \mathbb{C}^{M_RM_T\times G^2},
\label{Phi}
\end{equation}
where $\bm{\Phi}$ is called the measurement matrix and the covariance matrix of $\tilde{\bm{n}}^k$ is $\bm{R}_{\tilde{\bm{n}}}=\bm{I}_{M_T}\otimes \bm{R}_{\tilde{\bm{n}}_p}$. 

Combining the estimation process of all $K$ subchannels together, we have
\begin{equation}
\bm{Y}=\bm{\Phi X}+\tilde{\bm{N}},
\label{signal_model}
\end{equation}
where $\bm{Y}=[\bm{y}^1,\bm{y}^2,\cdots,\bm{y}^K]\in \mathbb{C}^{M_RM_T\times K}$, $\bm{X}=[\bm{x}^1,\bm{x}^2,\cdots,\bm{x}^K]\in \mathbb{C}^{G^2\times K}$, and $\tilde{\bm{N}}=[\tilde{\bm{n}}^1,\tilde{\bm{n}}^2,\\ \cdots,\tilde{\bm{n}}^K]\in \mathbb{C}^{M_RM_T\times K}$. For uplink channel estimation, we aim to design a matrix $\bm{\Phi}$ for efficient channel measurement and an estimation algorithm to accurately recover $\bm{H}=[\bm{h}^1,\bm{h}^2,\cdots,\bm{h}^K]\in\mathbb{C}^{N_RN_T\times K}$ based on $\bm{Y}$, $\bm{\Phi}$, and $\bm{R}_{\tilde{\bm{n}}}$\footnote{The original noise variance $\sigma^2$ can be calculated offline based on noise power spectrum density and bandwidth, or be estimated online periodically. Then, the effective noise covariance matrix can be constructed based on $\sigma^2$ and the designed $\bm{\Phi}$.}. Besides, due to the phase-shifter-based implementation, beam elements are restricted to have constant modulus, i.e., $|[\bm{W}]_{i,j}|=\frac{1}{\sqrt{N_R}},|[\bm{F}]_{i,j}|=\frac{1}{\sqrt{N_T}},\forall i,j$. 



\section{Conventional SBL-Based Channel Estimation Approaches}
Before diving into the proposed DL-based approach, we will first introduce the conventional SBL-based approaches in this section to clarify the underlying principles. 

As one of the most classic CS algorithms, SBL usually shows better performance than competitors like OMP thanks to the effective exploitation of signal sparsity. Consider the standard CS problem in Equation (\ref{signal_model_vector}). To estimate a certain subchannel in SBL, the sparse angular channel vector is modeled as Gaussian distribution with a diagonal covariance matrix such that
\begin{equation}
\bm{x}^k\sim \mathcal{CN}(\bm{0},\bm{R}_{\bm{x}^k}),
\end{equation}
\begin{equation}
\bm{R}_{\bm{x}^k}=\mathrm{diag}(\bm{\gamma}^k)\in \mathbb{R}^{G^2\times G^2},
\end{equation}
where $\bm{\gamma}^k=[\gamma_1^k,\gamma_2^k,\cdots,\gamma_{G^2}^k]$ denotes the variance parameters of the elements of $\bm{x}^k$. After initializing $\bm{R}_{\bm{x}^k}$ as $\bm{I}_{G^2}$, the following expectation-maximization (EM) steps are executed iteratively until convergence\cite{sbl}, and the eventual posterior mean is regarded as the estimation of $\bm{x}^k$:
\begin{itemize}
\item {\bfseries E-step:} Evaluate the posterior mean and the posterior covariance: 
\begin{equation}
\bm{\mu}_{\bm{x}^k}=\bm{R}_{\bm{x}^k}\bm{\Phi}^H(\bm{\Phi}\bm{R}_{\bm{x}^k}\bm{\Phi}^H+\bm{R}_{\tilde{\bm{n}}})^{-1}\bm{y}^k,
\label{a_posterior_mean}
\end{equation}
\begin{equation}
\bm{\Omega}_{\bm{x}^k}=\bm{R}_{\bm{x}^k}-\bm{R}_{\bm{x}^k}\bm{\Phi}^H(\bm{\Phi}\bm{R}_{\bm{x}^k}\bm{\Phi}^H+\bm{R}_{\tilde{\bm{n}}})^{-1}\bm{\Phi}\bm{R}_{\bm{x}^k}.
\end{equation}
\item {\bfseries M-step:} Update variance parameters:
\begin{equation}
\bm{\gamma}^k=|\bm{\mu}_{\bm{x}^k}|^2+\mathrm{diag}(\bm{\Omega}_{\bm{x}^k}).
\label{update_gamma}
\end{equation}
\end{itemize}

The original SBL algorithm is designed for sparse vectors whose elements are independent with each other. To exploit various channel sparsity structures, the following variants of SBL can be used for performance improvement:
\begin{itemize}
\item{\bfseries M-SBL:} To exploit the common sparsity structure in the frequency domain, instead of running SBL for $K$ times to estimate different subchannels separately, M-SBL estimates all subchannels together, where ``M" stands for MMV\cite{m_sbl}. In M-SBL, $K$ subchannel vectors $\bm{x}^1,\bm{x}^2,\cdots,\bm{x}^K$ share common $\bm{\gamma}$ and $\bm{\Omega_x}$. Therefore, Equation (\ref{update_gamma}) is modified to
\begin{equation}
\bm{\gamma}=\frac{\sum_{k=1}^K|\bm{\mu}_{\bm{x}^k}|^2}{K}+\mathrm{diag}(\bm{\Omega_x}).
\label{update_gamma_mmv}
\end{equation}

\item{\bfseries PC-SBL:} To exploit the block sparsity structure in the angular domain, variance parameters of nearby elements in PC-SBL are designed to be entangled with each other, where ``PC" stands for pattern-coupled\cite{pc_sbl_2d}. First, $\bm{\gamma}^k,|\bm{\mu}_{\bm{x}^k}|^2$, and $\mathrm{diag}(\bm{\Omega}_{\bm{x}^k})$ are arranged into $(G,G)$-dimensional matrices $\bm{\Gamma}^k$, $\bm{F}_1^k$, and $\bm{F}_2^k$ in the order of angular grids' sine values. Introducing the $(G,G)$-dimensional auxiliary variable matrix $\bm{A}^k$, we have $[\bm{\Gamma}]_{i,j}^k=([\bm{A}^k]_{i,j}+\beta([\bm{A}^k]_{i-1,j}+[\bm{A}^k]_{i+1,j}+[\bm{A}^k]_{i,j-1}+[\bm{A}^k]_{i,j+1}))^{-1},\forall i,j$, where $\beta$ denotes the coupling parameter and $[\bm{A}^k]_{0.}, [\bm{A}^k]_{.0}$ are always $0$. To update $\bm{A}^k$, we have $[\bm{A}^k]_{i,j}=\frac{a}{0.5\omega_{i,j}+b},\forall i,j$, where
\begin{equation}
\begin{aligned}
\omega_{i,j}=[\bm{F}_1^k]_{i,j}+[\bm{F}_2^k]_{i,j}+&\beta([\bm{F}_1^k]_{i-1,j}+[\bm{F}_2^k]_{i-1,j}+[\bm{F}_1^k]_{i+1,j}+[\bm{F}_2^k]_{i+1,j}\\
&+[\bm{F}_1^k]_{i,j-1}+[\bm{F}_2^k]_{i,j-1}+[\bm{F}_1^k]_{i,j+1}+[\bm{F}_2^k]_{i,j+1}),
\label{pc_update}
\end{aligned}
\end{equation}
$a$ is the shape parameter and $b$ is the inverse scale parameter of the Gamma hyperprior. $[\bm{F}_1]_{0.},[\bm{F}_1]_{.0},[\bm{F}_2]_{0.},[\bm{F}_2]_{.0}$ are $0$. When $\beta=0,a=0.5,b=0$, PC-SBL reduces to SBL.

\item{\bfseries M-PC-SBL:} To exploit both sparsity structures simultaneously, we can naturally combine M-SBL and PC-SBL. Similar to M-SBL, common $\bm{\gamma}$ and $\bm{\Omega_x}$ are shared by $K$ subchannels, and $\bm{F}_1^k$ in Equation (\ref{pc_update}) is replaced by $\frac{\sum_{k=1}^K\bm{F}_1^k}{K}$.
\end{itemize}

Nevertheless, the above SBL variants have limitations in the considered problem. Specifically, in PC-SBL, hyperparameters need to be carefully selected and the variance update rule is suboptimal\cite{pc_sbl_2d_ce}. Besides, the circularity of block sparsity is not considered. For M-SBL, its gain will be partially offset by beam squint since the common sparsity among subchannels becomes imperfect. M-PC-SBL may have the opposite effect with beam squint, although the operation of averaging over all subchannels originally aims to provide more accurate local entanglement information.

\section{DL-Based Channel Estimation Approach} 
To overcome the shortcomings of the conventional SBL-based approaches, we propose a DL-based channel estimation approach in this section. Next, the overall framework, network architecture design, training details, and complexity analysis will be elaborated sequentially.

\subsection{Framework of SBL Unfolding}
\begin{figure}[htb!]
\centering
\includegraphics[width=0.8\textwidth]{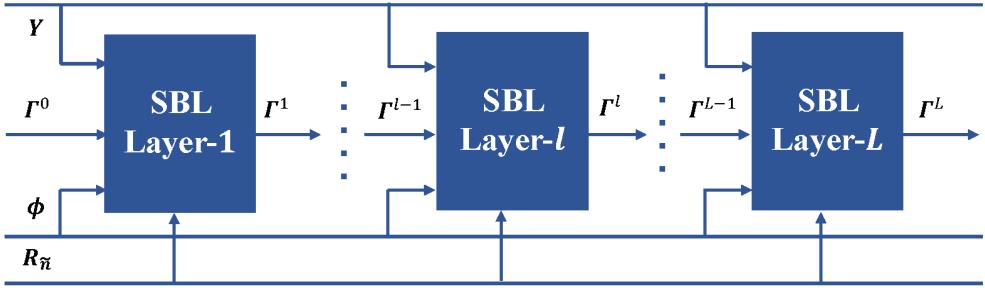}
\caption{The framework of an $L$-layer SBLNet.}
\label{framework}
\end{figure}

In general, the SBL algorithm with multiple iterations is unfolded into a DNN with cascaded layers, called the ``SBLNet", and the update rule of variance parameters in each SBL layer of the SBLNet is realized by a dedicated DNN. To facilitate the exploitation of the imperfect common sparsity structure in the frequency domain, a unique variance parameter is assigned to each element of $\bm{X}$ unlike M-SBL that uses shared variance parameters among different subchannels. Denote the variance parameter matrix in each SBL layer as $\bm{\Gamma}=[\bm{\gamma}_1,\bm{\gamma}_2,\cdots,\bm{\gamma}_k]\in \mathbb{R}^{G^2\times K}$. As illustrated in Fig. \ref{framework}, in an $L$-layer SBLNet, the input of the $l$-th SBL layer is $\bm{Y}$, $\bm{\Phi}$, $\bm{R}_{\tilde{\bm{n}}}$, and $\bm{\Gamma}^{l-1}$, and the output is the updated $\bm{\Gamma}^{l}$. The all-one $\bm{\Gamma}^0$ is input to the first SBL layer as the initialization of variance parameters, and the angular domain channel estimation $\hat{\bm{X}}$ can be readily obtained according to Equation (\ref{a_posterior_mean}) after $\bm{\Gamma}^L$ is output by the last SBL layer.

Stack all subchannels' posterior mean vectors' squared modulus and posterior covariance matrices' diagonal vectors to obtain two feature matrices $\bm{F}_{\mathrm{1}}=[|\bm{\mu}_{\bm{x}^1}|^2,|\bm{\mu}_{\bm{x}^2}|^2,\cdots,|\bm{\mu}_{\bm{x}^K}|^2]\in\mathbb{R}^{G^2\times K}$ and $\bm{F}_{\mathrm{2}}=[\mathrm{diag}(\bm{\Omega}_{\bm{x}^1}),\mathrm{diag}(\bm{\Omega}_{\bm{x}^2}),\cdots,\mathrm{diag}(\bm{\Omega}_{\bm{x}^K})]\in\mathbb{R}^{G^2\times K}$. Then, the optimal variance parameter update rule in each SBL layer, although may not be mathematically derivable under complex circumstances, should be a mapping function in the form of $\bm{\Gamma}_{\mathrm{updated}}=g(\bm{F}_{\mathrm{1}},\bm{F}_{\mathrm{2}})$, which can be parameterized by a DNN\footnote{In this paper, we train different parameters for DNNs in different SBL layers to achieve high performance. Nevertheless, DNNs in all SBL layers can also share the same parameters to improve parameter efficiency at the cost of performance degradation.} as $g_{opt}(\cdot;\bm{\Theta}_{opt})$ and the optimal network parameter set $\bm{\Theta}_{opt}$ can be learned through training\cite{l_sbl}.

As for the measurement matrix, instead of using hand-crafted matrices for $\bm{W}$ and $\bm{F}$ such as those consisting of random phases or Zadoff-Chu sequences\cite{s_sbl}, we jointly optimize them with the channel estimator. By modelling the phases of phase shifters as trainable parameters, they can be updated together with the weights of all SBL layers' DNNs during training\cite{mmv_amp,location_aided}. After training, the learned $\bm{W}$ and $\bm{F}$ can be deployed at the BS side and user side, respectively.

\subsection{Network Architecture}
Although DNNs are universal approximators theoretically, the specific architecture is critical to the performance of a DNN in practice. First of all, as illustrated in Fig. \ref{channel}, the circular block sparsity exists in the two-dimensional AoA-AoD space. Therefore, similar reshaping as in PC-SBL is performed to facilitate the update of variance parameters. Specifically, in the $l$-th SBL layer, the $(G^2,K)$-dimensional feature matrices $\bm{F}_1^{l-1}$ and $\bm{F}_2^{l-1}$ are converted to $(G,G,K)$-dimensional tensors where AoA and AoD are two separate dimensions, and the $(G,G,K)$-dimensional tensor of variance parameters is updated by the DNN, which is then converted back to the $(G^2,K)$-dimensional matrix, i.e., $\bm{\Gamma}^{l}$, for the $(l+1)$-th SBL layer's feature computation. The following key designs are adopted in the proposed network architecture to exploit channel sparsity structures:
\begin{itemize}
\item {\bf 3D Convolution:} Both the block sparsity in the angular domain and the imperfect common sparsity in the frequency domain result in local correlation in the input feature tensor, which can be well captured by the convolution operation. We use 3D convolution that is often used to process video data\cite{Conv3D}, where filters slide over the input tensor along its three dimensions, namely AoA, AoD, and subcarrier, to obtain the output tensor. Notice that, compared to the fully-connected (FC) layer used in \cite{l_sbl}, convolutional layer not only has better performance, but also has much lower complexity in large scale systems where the 
number of neuron connections of the FC layer explodes.
\item{\bf Circular Padding:} In CNNs, padding is necessary to keep the dimensions of feature maps unchanged after convolution. Instead of the common zero padding (ZP) used in most CNN-based channel estimation works\cite{cnn_refine,location_aided}, to deal with the circularity of block sparsity in the angular domain, we use circular padding (CP) that is often used to process panoramic images\cite{circular_padding}, where the top (and left) part is copied and concatenated to the bottom (and right) of the feature map, and vise versa. Notice that, ZP is still used in the subcarrier dimension.
\item{\bf Position Features Input}: One of the key features of CNN is that the convolution filters are shared and position-independent processing are executed when filters slide over the input feature map. However, unlike natural images, in the considered problem, the local correlation patterns in the feature tensor are naturally position-dependent. For instance, the squint directions in different AoA-AoD sections are naturally different, as illustrated by the directions of arrows in Fig. \ref{channel}. To realize position-dependent dynamic processing, we explicitly inform convolution filters of their current position\cite{coord_conv} by incorporating the sine values of the covered angular grids' AoAs and AoDs as additional features. Denote the  $(G,G,K,2)$-dimensional position features as $\bm{F}_3$, we have $[\bm{F}_3]_{i,j,k,0}=sin(\phi_i)$ and $[\bm{F}_3]_{i,j,k,1}=sin(\phi_j),\forall i,j,k$. 
\end{itemize}

\begin{figure}[htb!]
\centering
\includegraphics[width=1\textwidth]{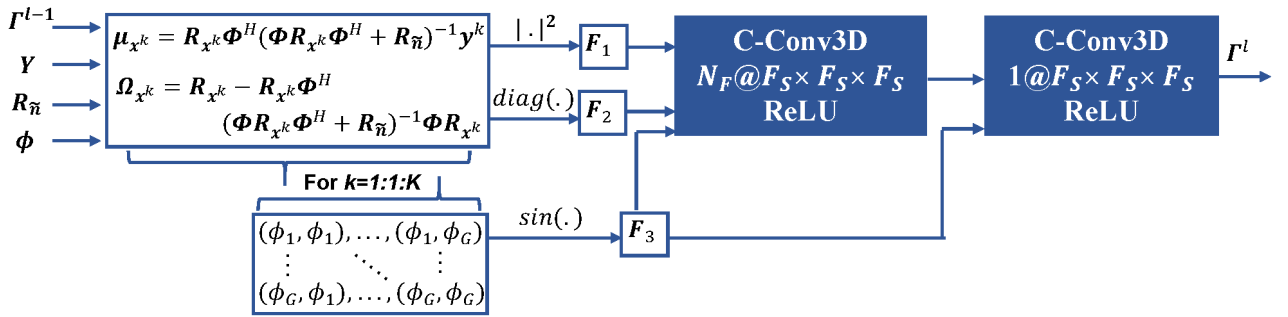}
\caption{Network architecture of the DNN in the $l$-th SBL layer.}
\label{network}
\end{figure}

The detailed architecture of the DNN in the $l$-th SBL layer is illustrated in Fig. \ref{network}. The $(G,G,K,4)$-dimensional input feature map is obtained by stacking $\bm{F}_1,\bm{F}_2,\bm{F}_3$ along the last dimension. We use two 3D convolutional layers with CP (C-Conv3D), where the first layer has $N_F$ filters to extract intermediate features and the second layer has only one filter to output the updated $\bm{\Gamma}^{l}$. The filter size is $F_S$. ReLU activation is used after both layers, where the first ReLU introduces nonlinearity and the second ReLU outputs non-negative predictions. 

\subsection{Network Training}
For network training, we generate $10,000$ data samples in total, and split them into a training set, a validation set, and a testing set with a ratio of 8:1:1. Adam is used as the optimizer. The mean-squared error (MSE) is used as the loss function, i.e., $\mathrm{Loss} = \frac{1}{SKN_TN_R}(||\bm{H}_s-\hat{\bm{H}}_s||_F^2)$, where the number of samples in a mini-batch, $S$, is set to 16 in experiments. Three-stage training strategy is used. Specifically, in the first stage, weights of convolution filters are initialized using Xavier and trained with different random $\bm{W}$ and $\bm{F}$ until convergence. In the second stage, the convolution filters are frozen and only the phases of $\bm{W}$ and $\bm{F}$ are trained to obtain the optimal measurement matrix. In the first two stages, learning rate is initialized as $10^{-4}$ and decays with a factor of 10 and a patience of 2 epochs to accelerate training, and early stopping with a patience of 3 epochs is used to avoid overfitting. In the third stage, the entire network is fine-tuned with a small learning rate $10^{-5}$, and early stopping is also used.

\subsection{Complexity Analysis}
The number of real floating operations (FLOPs) is calculated to compare the complexity of different algorithms. For algorithms in the SBL family, the FLOPs per iteration of complexity-dominating operations is $16KG^2(M_RM_T)^2$. As for the proposed SBLNet, the extra real FLOPs per iteration introduced by the DNN is $2(5N_F+2)F_S^3KG^2$. We only demonstrate the online prediction complexity here since the offline training does not happen very often and the BS is assumed to have sufficient memory and computation resources. On the NVIDIA GeForce RTX 3090 GPU, network training usually takes about several hours. With typical system and network parameters, the increase of complexity per iteration of SBLNet is marginal while the number of iterations can be dramatically reduced compared to SBL, as will be shown later in simulation. Therefore, the overall complexity of SBLNet is lower than SBL, making it appealing for practical applications.

\section{Extension to the Multi-Block Case}
\label{multi_block}
So far we have been discussing channel estimation in a single block. In this section, we further extend the proposed approach to the multi-block case and exploit channel correlation in the time domain for further performance improvement.

\subsection{Time-Selective Channel Model}
Since the angles and delays of channel paths change relatively slow, it is reasonable to assume that they are constant for several consecutive blocks\cite{sbl_ce,cnn_refine}. So, the $k$-th subchannel in the $n$-th block can be expressed as
\begin{equation}
\bm{H}^k[n]=\sqrt{\frac{N_TN_R}{N_cN_p}}\sum_{i=1}^{N_c}\sum_{j=1}^{N_p}\alpha_{i,j}[n]e^{-j2\pi\tau_{i}f_s\frac{k}{K}}\bm{a}_R(\theta^R_{i,j},k)\bm{a}_T(\theta^T_{i,j},k)^H,
\end{equation}
where the evolution of path gains in time follows the first order autoregressive process as 
\begin{equation}
\alpha_{i,j}[n]=\rho \alpha_{i,j}[n-1]+\sqrt{1-\rho^2}w_{i,j}[n],\forall i,j,
\end{equation}
and the temporal correlation coefficient can be computed according to the Jakes' model as $\rho=J_0(2\pi f_D\triangle_t)$, where $J_0$ is the zeroth-order Bessel function of the first kind, $\triangle_t$ is the block length, and $f_D=vf_c/c$ is the maximum Doppler frequency with $v$ and $c$ denoting the speed of user and light, respectively. Besides, $w_{i,j}[n]\sim\mathcal{CN}(0,1)$ denotes the innovation noise. The received signal in the $n$-th block can be expressed as 
\begin{equation}
\bm{Y}[n]=\bm{\Phi}[n]\bm{X}[n]+\tilde{\bm{N}}[n],
\end{equation}
where the measurement matrix here can also be time-varying rather than long-term fixed.

\subsection{Problem Reformulation}
To exploit channel correlation in the time domain, one method is to estimate the channels of multiple blocks together, e.g., by using algorithms like M-SBL and T-SBL\cite{sbl_ce}. Nevertheless, in this way, the channel estimations of previous blocks are not available until the last block's pilots arrive, which is not practical since the CSI needs to be used for signal detection and precoding in each block immediately. Therefore, a more proper way is to estimate online and exploit previous blocks' estimation results to aid current block's estimation\cite{sbl_ce}.

In this paper, we only exploit one previous block for simplicity while exploiting several previous blocks is totally feasible. We use the modulus of angular domain channel estimation, $|\hat{\bm{X}}[n-1]|$, as the time features since they reflect the short-term channel sparsity patterns. Then, the time features are exploited in two aspects, namely facilitating the update of variance parameters and designing the adaptive measurement matrix. The former is based on that consecutive blocks' angular domain channels have similar positions of non-zero elements while the latter is based on that the measurement efficiency of the measurement matrices can be improved by focusing energy on angular grids with high confidence\cite{location_aided}. 

\subsection{Modifications to the Network Architecture}
\begin{figure}[htb!]
\centering
\includegraphics[width=1\textwidth]{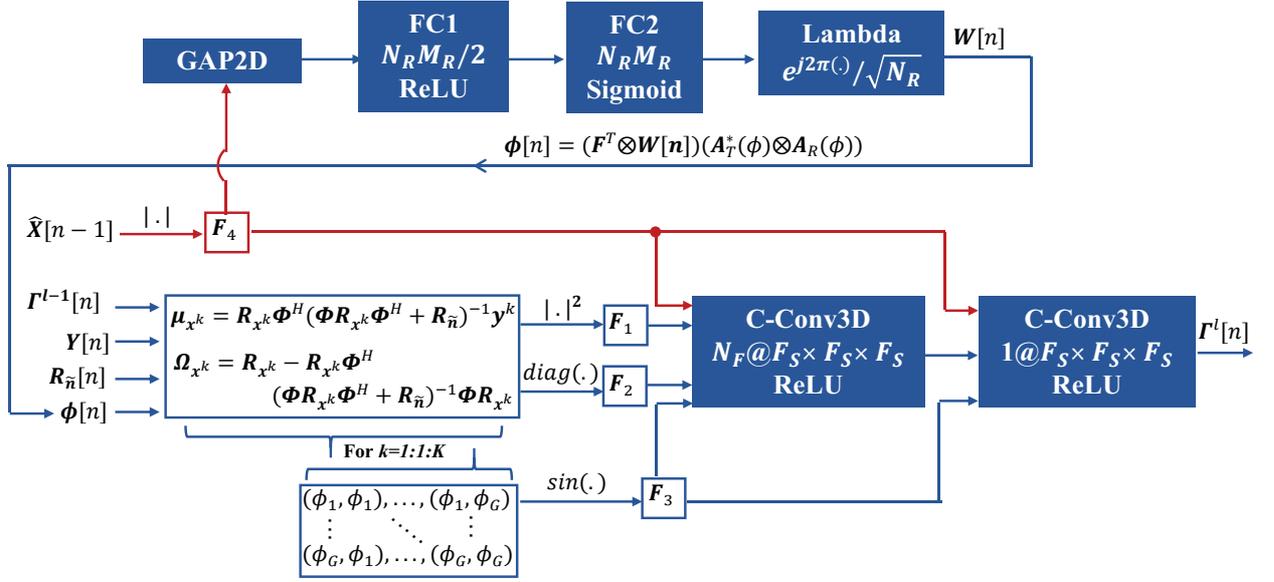}
\caption{Network architecture adopted in the multi-block case.}
\label{network_multi}
\end{figure}

To realize the multi-block channel estimation, the network architecture illustrated in Fig. \ref{network_multi} is adopted. Compared to Fig. \ref{network}, the first modification is to incorporate the $(G,G,K,1)$-dimensional time features $|\bm{\hat{X}}[n-1]|$, i.e., $\bm{F}_4$, as an extra part of each DNN layer's input in each SBL layer of the SBLNet. The second modification is that another dedicated DNN is used for measurement matrix prediction. Notice that only the receive combining matrix $\bm{W}[n]$ is adaptively predicted and the transmit beamforming matrix $\bm{F}$ is still fixed since it is deployed at the user side while previous blocks' estimation results are only available at the BS in TDD systems. Specifically, a two-dimensional global average pooling (GAP2D) layer is used first to average out the dimensions of subcarrier and AoD in $\bm{F}_4$ since all subcarriers share a common measurement matrix and $\bm{W}$ at the user side is for AoA sensing. Then, based on the $G$-dimensional squeezed feature vector, two FC layers are used to predict the phases of $\bm{W}[n]$, where the first FC layer has $\frac{N_RM_R}{2}$ neurons with a ReLU activation and the second FC layer has $N_RM_R$ neurons with a Sigmoid activation. Eventually, a Lambda layer is used to scale the predictions to phases between $[0,2\pi]$, convert the phases to complex numbers, and perform modulus normalization. With the predicted $\bm{W}[n]$ and the optimized $\bm{F}$ in the single-block case, $\bm{\Phi}[n]$ can be readily computed. 

\subsection{Network Training and Complexity Analysis}
In the multi-block case, the modified SBLNet is trained alone first. For the initialization of weights of convolution filters in two C-Conv3D layers, those connecting to the newly added $\bm{F}_4$ are initialized as zero while the rest are copied from the previously trained model in the single-block case, so that the function of the modified SBLNet is exactly the same as the original one at the beginning of training\cite{dual_cnn}. After convergence, the convolution filters are frozen and the dedicated DNN for matrix prediction is introduced and trained to convergence, whose weights use Xavier initialization. Eventually, the entire network is fine-tuned with a small learning rate. Strategies including learning rate decay and early stopping are still used. 

Compared to the single-block approach, the extra complexity in each iteration of the multi-block approach comes from two aspects, namely the increased complexity brought by the additional time features in the DNN in each SBL layer of the SBLNet, and the complexity of the dedicated DNN for matrix prediction. The FLOPs of the former is $2(N_F+1)F_S^3KG^2$, and the FLOPs of the latter is $(G+N_RM_R)N_RM_R$. Compared to the single-block approach, the extra complexity is only marginal while decent performance gain can be achieved, as will be shown later in simulation.

\section{Simulation Results}
\label{simulation}
In this section, extensive simulation results are provided to validate the superiority of the proposed approaches in terms of performance, complexity, and robustness.
\subsection{Configuration}
Unless specified, the following parameters will be used as the typical setting in simulation: $N_T=32$, $N_R=32$, $N_R^{RF}=4$, $N_T^{RF}=1$, $N_c=3$, $N_p=10$, $\triangle\theta_i^R=\triangle\theta_i^T=5^\circ$, $f_c=28$ GHz, $v=1$ m/s, and $\triangle t=1$ ms so that the corresponding $\rho=0.916$. $f_s=4$ GHz, $K=8$, and SNR $=20$ dB. Normalized mean-squared error (NMSE) defined as $\mathbb{E}\{||\bm{H}-\hat{\bm{H}}||_F^2/||\bm{H}||_F^2\}$ is utilized to measure the channel estimation performance obtained by averaging over 300 Monte Carlo channel and noise realizations. The following algorithms are selected as baselines for comparison:
\begin{itemize}
\item {\bf SBL family\cite{sbl_ce,pc_sbl_2d_ce}:} SBL and PC-SBL are executed for $K$ times to estimate different subchannels separately while M-SBL and M-PC-SBL are executed once to estimate all subchannels together. Enough iterations (typically 100-300) are executed until performance converges and the best hyperparameters of PC-SBL and M-PC-SBL are searched.
\item {\bf LS with sufficient beams:} Use $M_R=N_R$ and $M_T=N_T$ beams to obtain the LS estimation by $\hat{\bm{H}}^k_{LS}=(\bm{W}^H)^{-1}\bm{Y}^k\bm{F}^{-1}$, which needs $\frac{N_TN_R}{M_TM_R}$ times overhead of all the other algorithms.
\item {\bf Data-driven DL\cite{fdd_autoencoder}:} Obtain coarse estimation with a FC layer first and use CNN for further refinement. Also, the measurement matrix is jointly learned with the channel estimator. In the considered problem, this method has similar performance as the method in \cite{attention} and better performance than the method in \cite{cnn_refine}.
\end{itemize}

\subsection{Determination of Network Architecture}
\begin{figure}[htb!]
\centering
\includegraphics[width=0.5\textwidth]{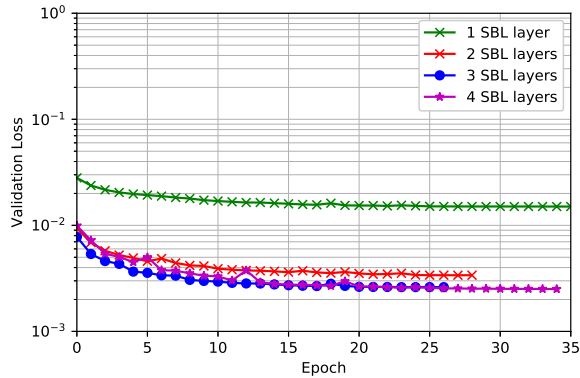}
\caption{Validation loss evolution during training with different numbers of SBL layers.}
\label{impact_of_SBL_layers}
\end{figure}

Network hyperparameters with the best performance-complexity tradeoff are determined by cross validation. Eventually, $L=3$, $N_F=8$, $F_S=5$ are used. As an example, the impact of the number of SBL layers $L$ is illustrated in Fig. \ref{impact_of_SBL_layers}, where $M_T=16,M_R=16,G=64$, and other parameters the same as the typical setting. We can see the network performance saturates with 3 SBL layers and more layers only improves performance slightly but leads to higher complexity. 

The effectiveness of each specific design adopted in the proposed network architecture is validated in Table \ref{alabation}, where $M_T=16,M_R=8,G=64$, and other parameters the same as the typical setting. As we can see, the network performance gradually improves from the vanilla version with more designs used, and eventually exceeds all conventional SBL-based algorithms. Specifically, the performance of Conv2D improves dramatically with the input of position features and becomes better than PC-SBL, which demonstrates the effective exploitation of the block sparsity structure in the angular domain. Then, Conv3D outperforms Conv2D due to further exploitation of the channel sparsity structure in the frequency domain. In contrast, M-PC-SBL has no performance gain over PC-SBL with best hyperparameters searched being $\beta=0,a=0.5,b=0$ since beam squint damages the common sparsity structure and makes the local entanglement information less accurate after averaging over all subchannels. Notice that, without beam squint, M-PC-SBL does outperform M-SBL and PC-SBL, which indicates that beam squint makes it hard to exploit both channel sparsity structures simultaneously using conventional SBL variants. At last, by using CP instead of ZP in the angular domain, channel paths near the edges of the channel images are better estimated, resulting in the eventual architecture with the best performance. 
\begin{table}[htb!]
\begin{tabular}{|c|c|}
\hline
Approach & NMSE\\ \hline
SBL & 0.1203 \\ \hline
PC-SBL & 0.1015 \\ \hline
M-SBL & 0.0803\\ \hline
M-PC-SBL & 0.0803\\ \hline
Conv2D & 0.3127\\ \hline
Conv2D + Position features & 0.0710 \\ \hline
Conv3D + Position features & 0.0429\\ \hline
{\bf Conv3D + CP + Position features} & {\bf 0.0333} \\ 
\hline
\end{tabular}
\centering
\caption{Effectiveness of the network architecture. Conv2D only performs convolution in the AoA-AoD space and different subchannels are estimated separately.} 
\label{alabation}
\end{table}

\subsection{Performance Analysis with A Single Block}
In this subsection, we analyze the performance of different algorithms in the single-block case. 

\subsubsection{Impacts of key system parameters}
\begin{figure}[htb!] 
\centering 
\vspace{-0.35cm} 
\subfigtopskip=2pt 
\subfigbottomskip=1pt 
\subfigcapskip=-3pt 
\subfigure[]{
\label{impact_of_grids}
\includegraphics[width=0.48\textwidth]{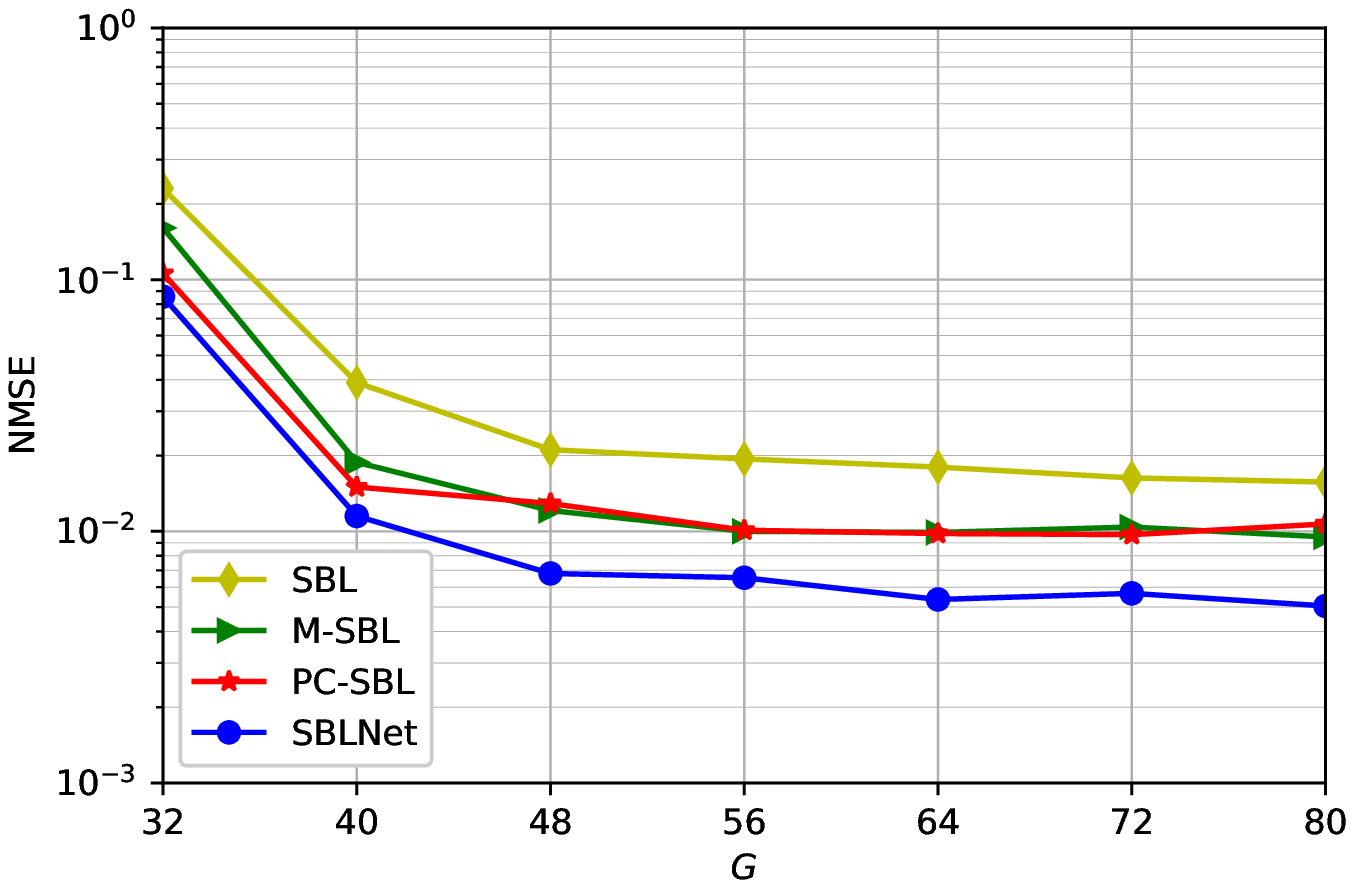}}
\subfigure[]{
\label{impact_of_beams}
\includegraphics[width=0.48\textwidth]{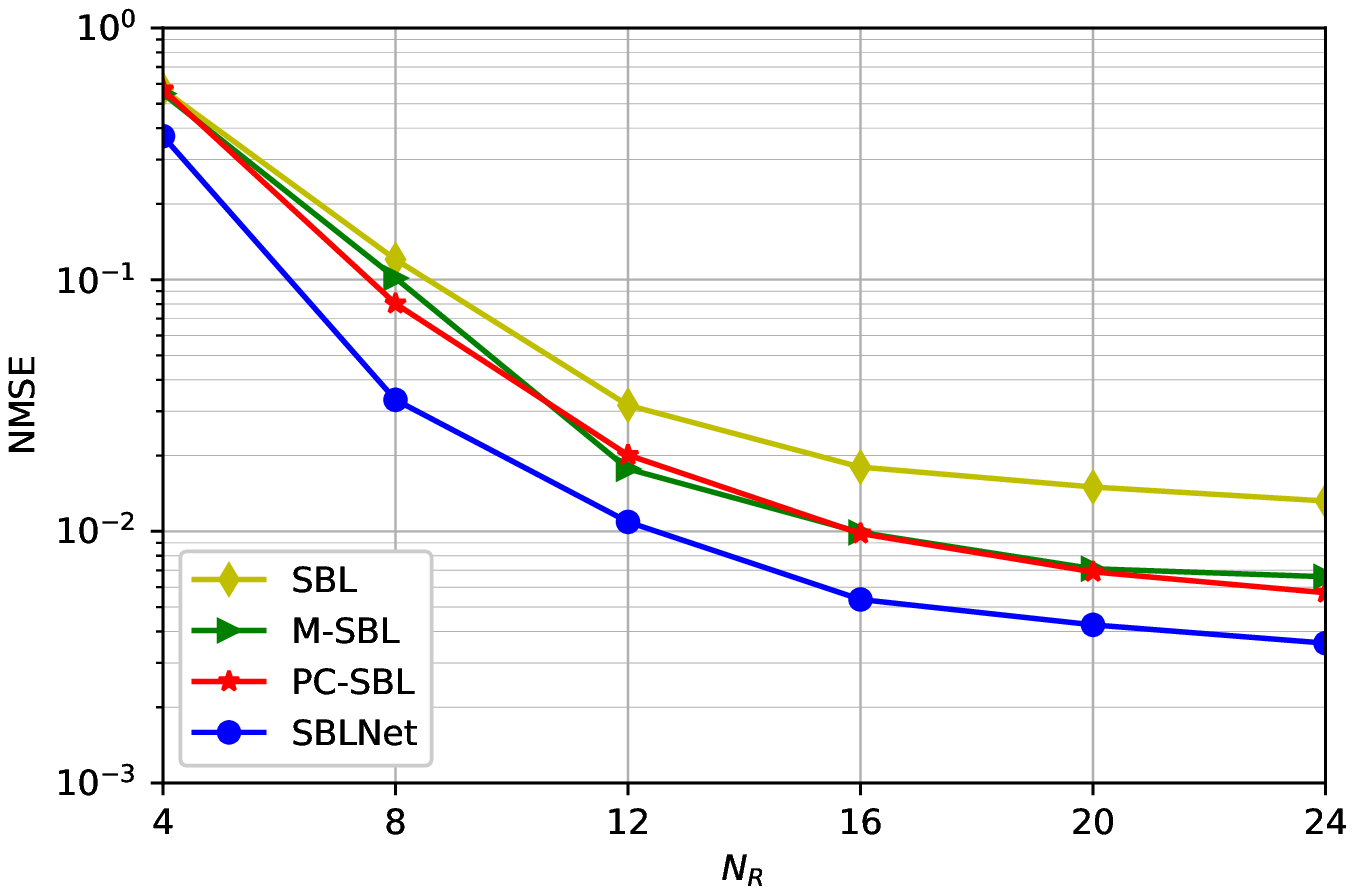}}
\caption{Impacts of the number of grids $G$ and the number of receive beams $N_R$.}
\label{impact_of_G_and_N_R}
\end{figure}

Random $\bm{W}$ and $\bm{F}$ are considered first. As illustrated in Fig. \ref{impact_of_grids}, the performance of all algorithms improves with the growth of $G$ at the beginning since dictionary matrices with higher angular resolutions can alleviate power leakage and enhance the sparsity of the angular domain channel\cite{off_grid_sbl_ce}. Then, when $G$ is large enough, the performance saturates. To keep the complexity low, $G=64$ will be used in the following experiments. Similarly, the impact of the number of receive beams is illustrated in Fig. \ref{impact_of_beams}. With more information brought by more beams, the performance gets better at the cost of higher estimation overhead and complexity. Similar trends can be observed when changing the number of transmit beams. To balance performance, complexity, and overhead, $N_R=M_R=16$ will be used in the following experiments. From Fig. \ref{impact_of_G_and_N_R}, we can see that the proposed SBLNet consistently outperforms three conventional SBL algorithms with various $G$ and $N_R$, which demonstrates its superiority.

\begin{figure}[htb!] 
\centering 
\vspace{-0.35cm} 
\subfigtopskip=2pt 
\subfigbottomskip=1pt 
\subfigcapskip=-3pt 
\subfigure[]{
\label{impact_of_snr}
\includegraphics[width=0.48\textwidth]{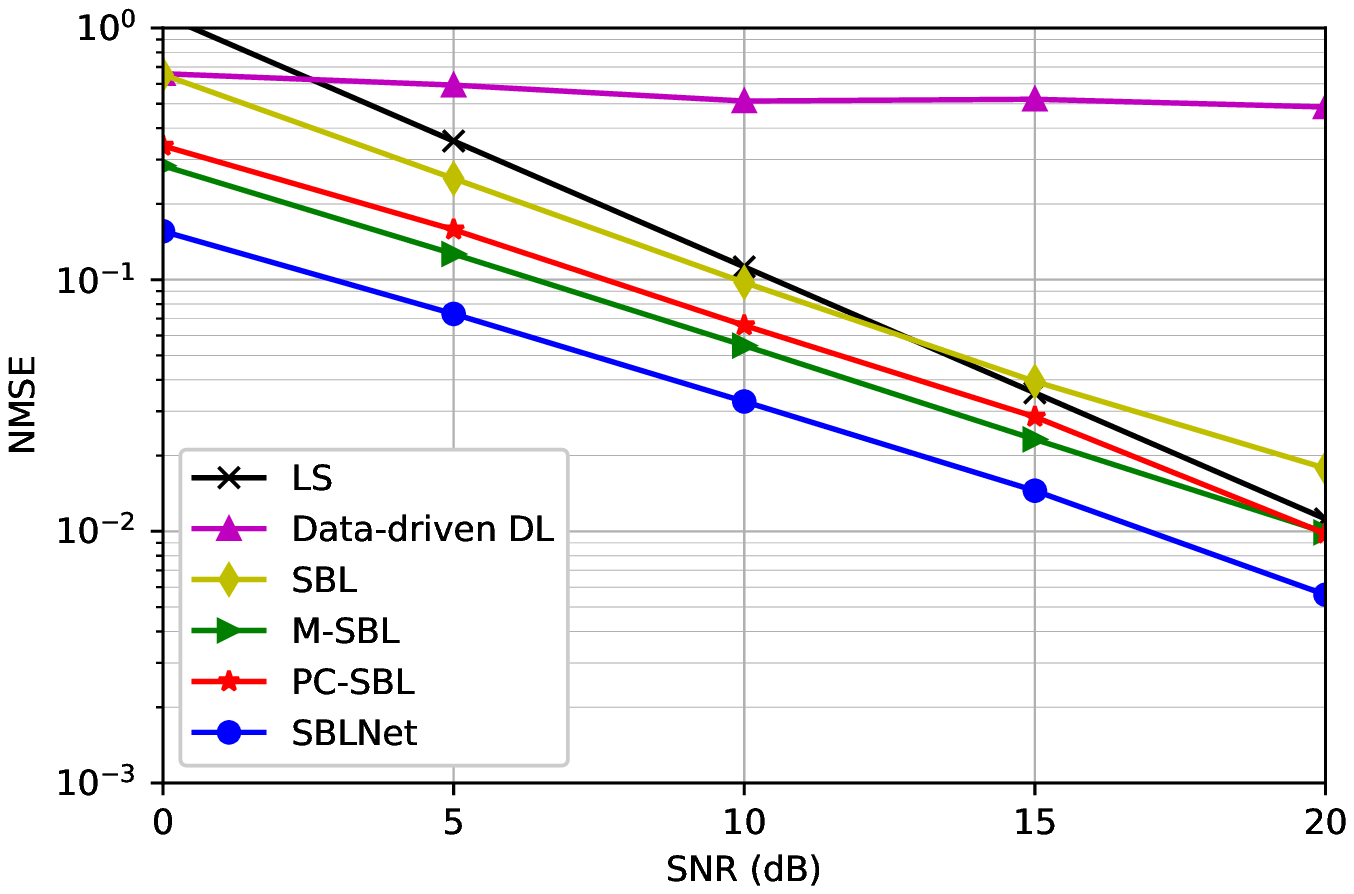}}
\subfigure[]{
\label{impact_of_measurement_matrix}
\includegraphics[width=0.48\textwidth]{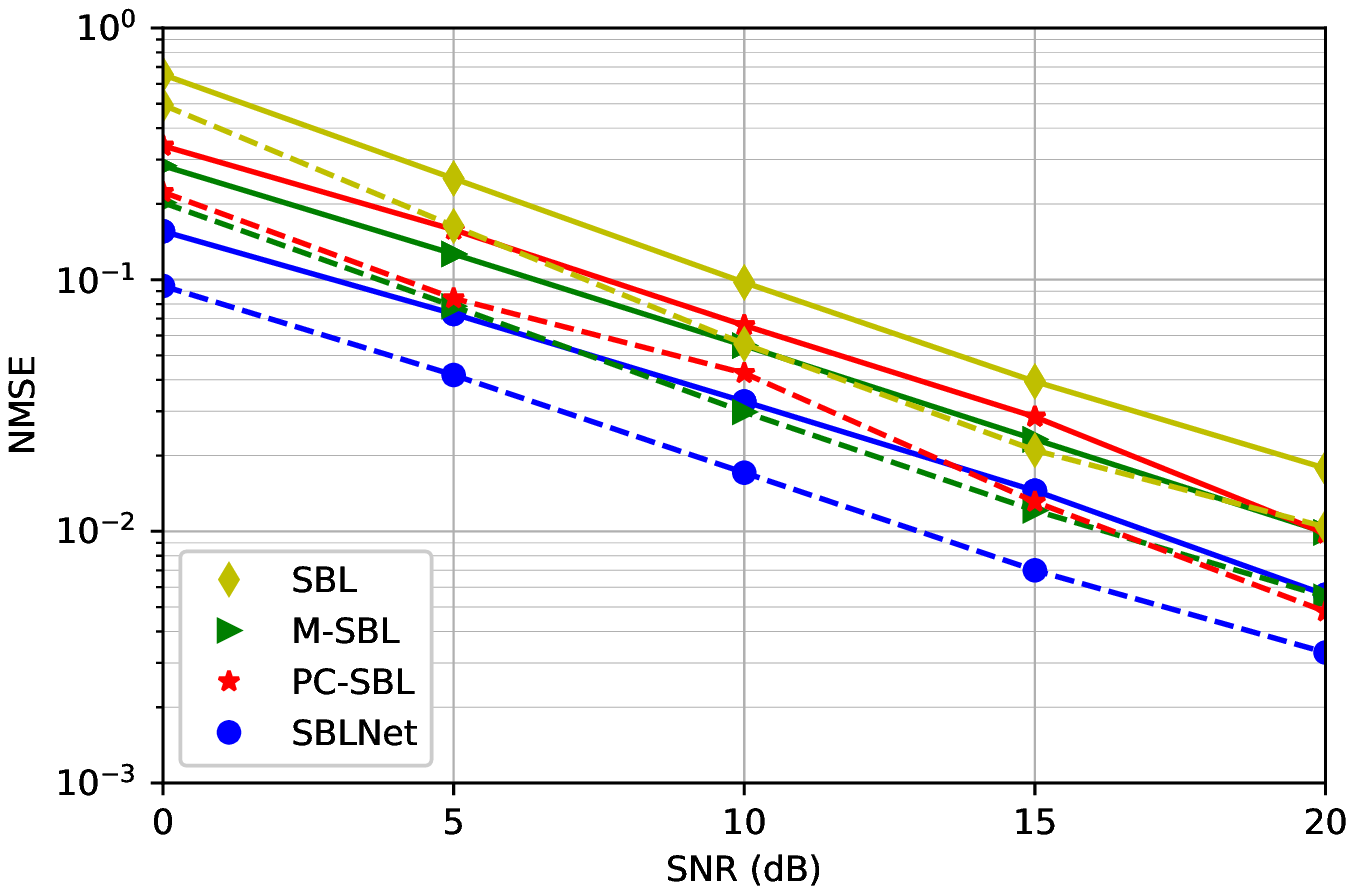}}
\caption{Impacts of SNR and the measurement matrix.}
\label{impact_of_G_and_N_R}
\end{figure}

The impact of SNR is illustrated in Fig. \ref{impact_of_snr}. As we can see, data-driven DL fails to work unless with very low SNR, which confirms the necessity of using model-driven DL-based approaches in the considered problem. Although fourfold estimation overhead is consumed to guarantee sufficient beams, the performance of LS is still inferior to SBLNet without considering channel sparsity. Both M-SBL and PC-SBL outperform the original SBL consistently in all SNR regimes, while their performance gains are limited because of the mismatch between the real complicated sparsity structures and the assumed ideal models. In contrast, SBLNet is able to learn arbitrary sparsity structures and achieves much more significant performance gain over SBL. Moreover, as shown in Fig. \ref{impact_of_measurement_matrix}, with the jointly optimized measurement matrix (dashed curves), all SBL-based methods achieve decent performance gain compared to their counterparts with random measurement matrices (solid curves). It is because the optimized matrix may have higher measurement efficiency and better properties for sparse signal recovery.  

\subsubsection{Generalization ability}
Strong generalization ability is critical to make a DL-based method practical. First of all, the proposed approach can naturally generalize to different system scales including $N_T,N_R,M_T,M_R$ thanks to the parameter sharing mechanism of CNN. Since convolution happens in the angular domain and frequency domain, the same network can be re-used as long as $G$ and $K$ are unchanged. On the other hand, parameters influencing the channel distribution may be time-varying in practice, such as SNR, the number of clusters, and the angular spread. 

\begin{figure}[htb!] 
\centering 
\vspace{-0.35cm} 
\subfigtopskip=2pt 
\subfigbottomskip=1pt 
\subfigcapskip=-3pt 
\subfigure[SNR generalization]{
\label{SNR_generalization}
\includegraphics[width=0.48\textwidth]{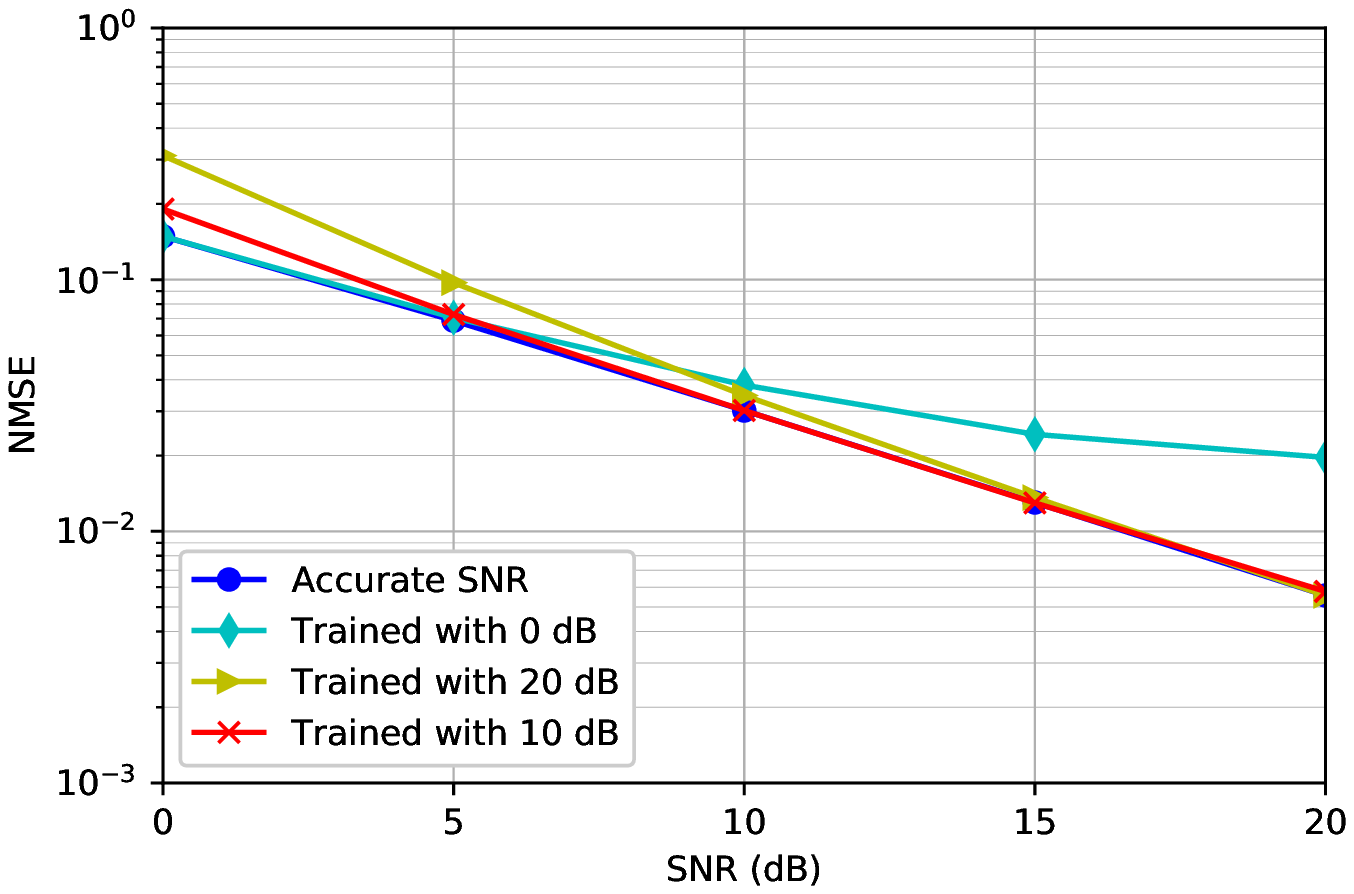}}
\subfigure[Cluster number and angular spread generalization]{
\label{channel_generalization}
\includegraphics[width=0.48\textwidth]{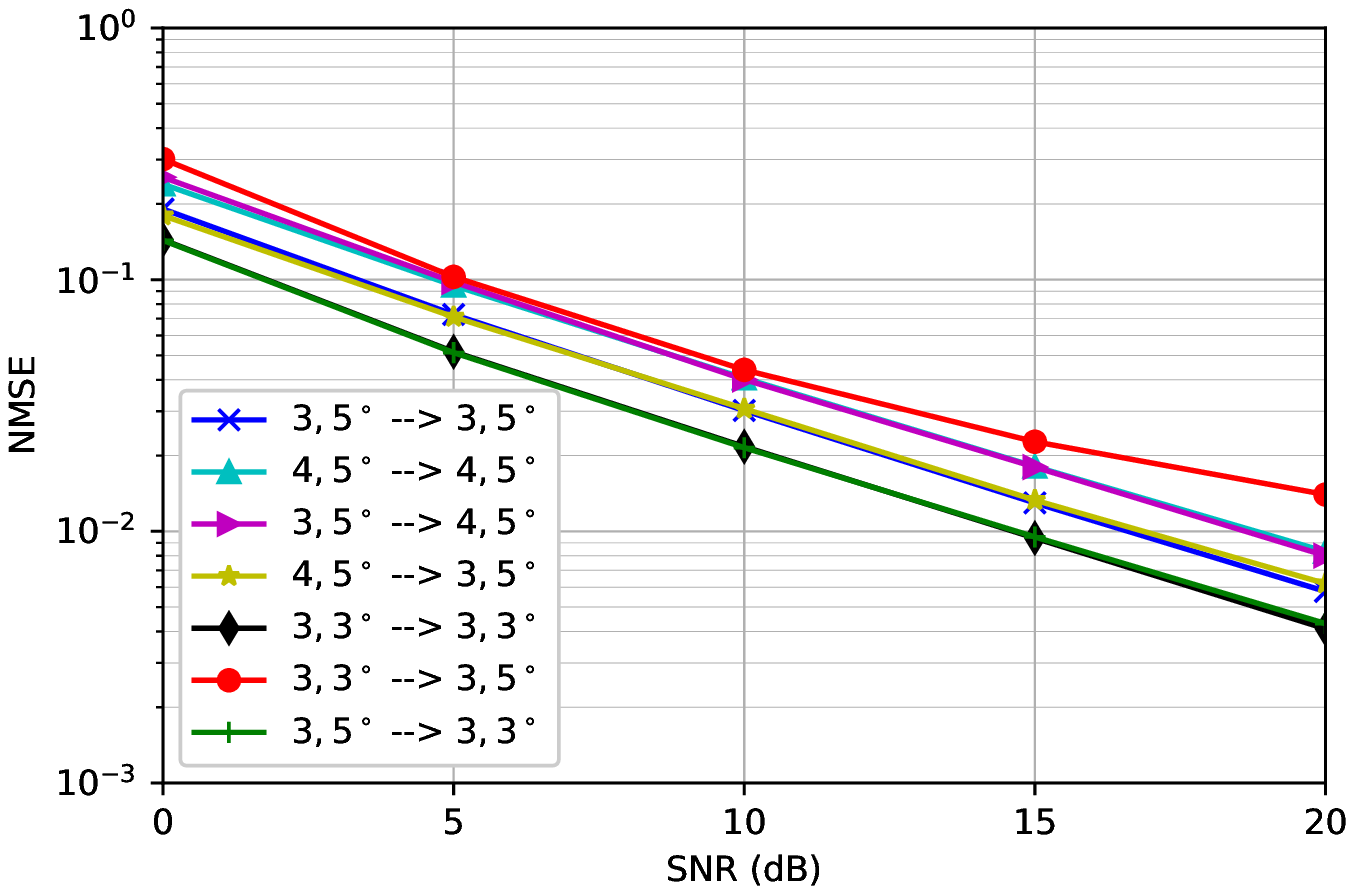}}
\caption{Generalization to key channel-related parameters. Random matrices are used. In the legends of the second subfigure, the two numbers before (or after) the arrow denote the number of clusters and the angular spread of the training set (or testing set), respectively.}
\label{generalization}
\end{figure}

Among several common methods to deal with SNR generalization\cite{location_aided,attention,SNR_attention}, we choose to test with the model trained by a moderate SNR point\cite{location_aided} since it is simple and can already achieve satisfactory generalization performance. As illustrated in Fig. \ref{SNR_generalization}, the performance of testing with the network trained by only 10 dB data is almost the same as testing with individual networks trained by accurate SNR points, except for slight performance degradation when SNR $=0$ dB. The trends are similar when the optimized measurement matrix is used.

The generalization to different numbers of clusters and angular spreads is illustrated in Fig. \ref{channel_generalization}. As we can see, the network trained with 2-cluster channel has no performance loss when testing 3-cluster channel and vice versa. Regarding the appearance of a cluster as a pattern, no matter how many clusters are there in the channel image, they can be captured when the convolution filters corresponding to this pattern slide over. As for angular spread, network trained with larger-angular-spread channel generalizes well to smaller-angular-spread channel since the latter can be regarded as the special case of the former. However, there is obvious performance degradation in the contrary case, as indicated by the red curve. Therefore, we should train the network with large-angular-spread channel in practice to ensure its robustness.

Thanks to the strong generalization ability, only a single network needs to be trained in practice to handle various situations, which dramatically enhances the practicality of the proposed approach.

\subsection{Performance Analysis with Multiple Blocks}
\begin{figure}[htb!]
\centering
\includegraphics[width=0.5\textwidth]{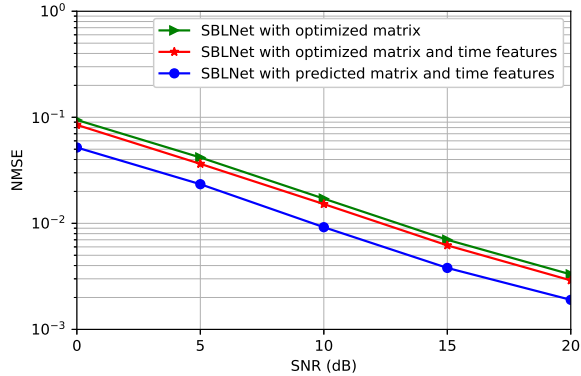}
\caption{Performance gain of the proposed multi-block approach.}
\label{impact_of_SNR_multi}
\end{figure}

In this subsection, we demonstrate the performance gain in the multiple-block case. From Fig. \ref{impact_of_SNR_multi}, with the jointly optimized measurement matrix, the incorporation of time features in SBLNet facilitates the update of variance parameters and decreases the NMSE slightly in all SNR regimes. Then, the performance further improves decently with the predicted measurement matrix, which validates the effectiveness of the proposed designs in the multi-block approach.

The generalization ability is also shown in Table \ref{multi_block_generalization}. On the one hand, when the trained network is tested with a different temporal correlation factor $\rho=0.707$, no performance degradation is observed since the most important information exploited in the proposed approach is the short-term locations of non-zero angular grids, which can always be reflected in time features regardless of the value of $\rho$. On the other hand, since it may be too strict to assume constant path angles in consecutive blocks, for each cluster, we add a disturbance following $\mathcal{U}[-3^\circ,3^\circ]$ to the AoAs and AoDs of all subpaths in the previous block\cite{angle_change} and test the robustness of the network. Again, no obvious performance degradation is observed, which demonstrates the strong generalization ability of the proposed approach in the multi-block case.
\begin{table}[htb!]
\begin{tabular}{|c|c|}
\hline
Testing Parameters & NMSE\\ \hline
$\rho=0.916$, constant angles & 0.0019\\ \hline
$\rho=0.707$, constant angles & 0.0019\\ \hline
$\rho=0.916$, changing angles & 0.0020\\ \hline
\end{tabular}
\centering
\caption{Generalization under the typical system setting in the multi-block case.} 
\label{multi_block_generalization}
\end{table}

\begin{figure}[htb!] 
\centering 
\vspace{-0.35cm} 
\subfigtopskip=2pt 
\subfigbottomskip=1pt 
\subfigcapskip=-3pt 
\subfigure[Example 1]{
\includegraphics[width=0.48\textwidth]{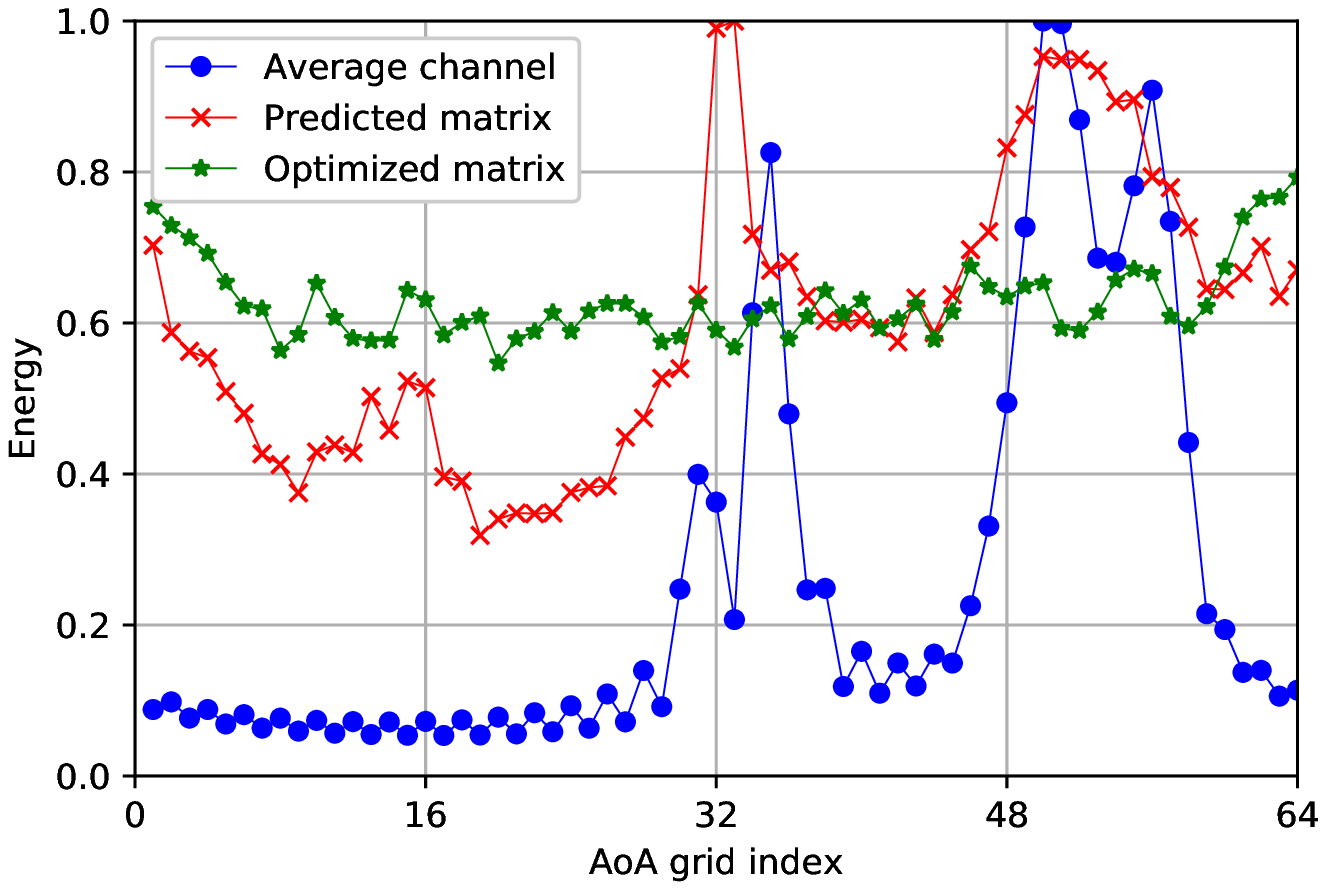}}
\subfigure[Example 2]{
\includegraphics[width=0.48\textwidth]{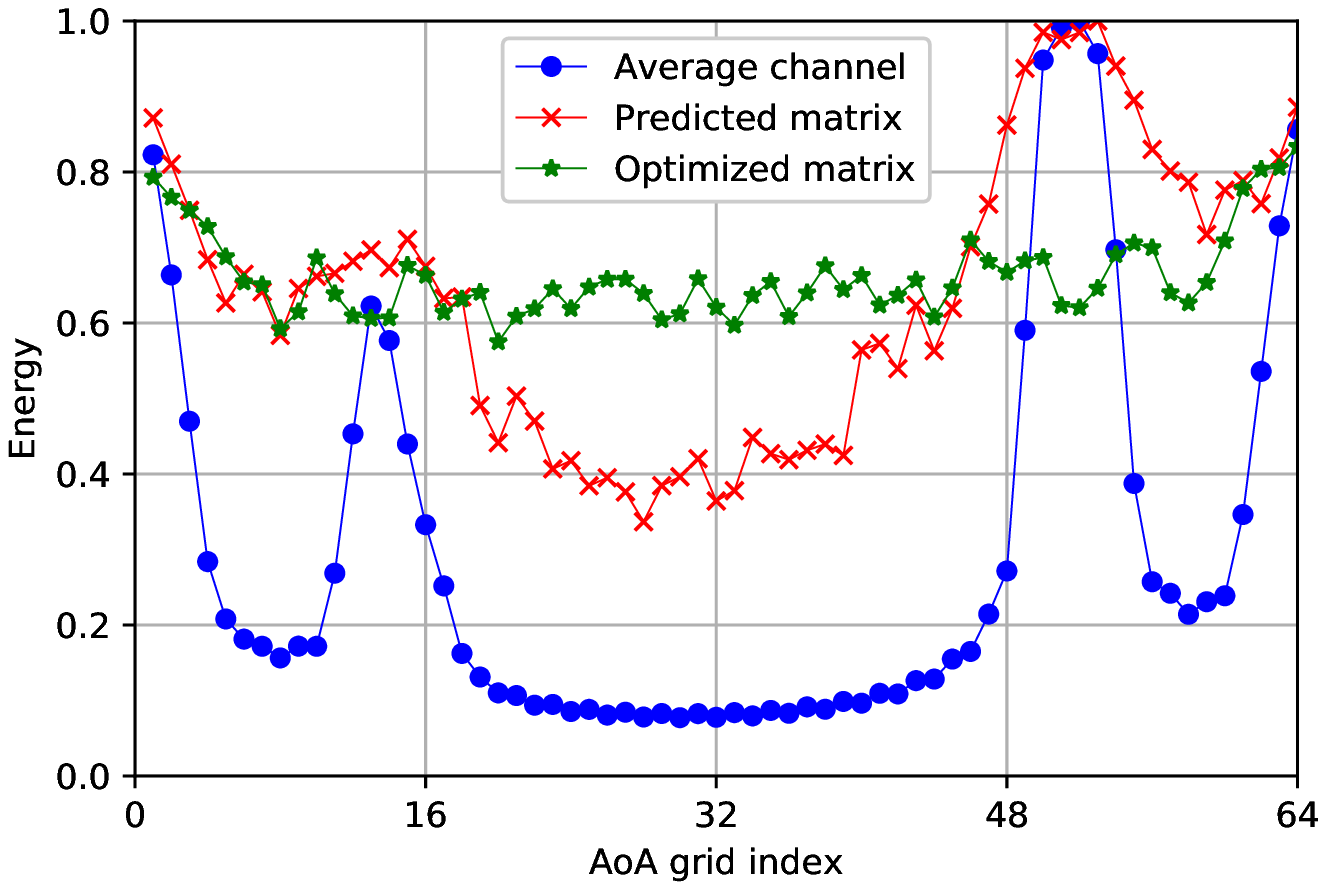}}
\caption{Examples of energy distributions of average channel and different measurement matrices on AoA grids. Proper scaling is executed for better display effect.}
\label{learned_matrix}
\end{figure}

To intuitively understand the superiority of the adaptively predicted measurement matrix, two examples of energy distributions of average channel and different measurement matrices on AoA grids are given in Fig. \ref{learned_matrix}. Specifically, in each subfigure, the $g$-th point on the blue curve is obtained by $\frac{1}{KG}\sum_{k=1}^{K}\sum_{g'=1}^G|[\bm{X}^k]_{gg'}|$, where the dimensions of AoD and subcarrier are averaged out while the $g$-th point on the green or red curve is obtained by $||[\bm{W}^H\bm{A}_R(\Phi)]_{.g}||$ or $||[\bm{W}[n]^H\bm{A}_R(\Phi)]_{.g}||$, i.e., the total measurement energies of all receive beams at the BS side. As we can see, the jointly optimized matrix in the single-block case has dispersive measurement energies throughout all grids since it is long-term fixed and has to consider all possible path angles fairly. In contrast, the adaptively predicted matrix has more focused measurement energies on those grids that are highly possible to have incoming paths, which are inferred from the time features. The perfect match with the short-term channel sparsity patterns well explains the performance gain of the predicted matrix. Notice that, although angular regions without obvious channel gains still correspond to moderate measurement energies of the predicted matrix due to limited degree of optimization freedom, it is actually beneficial to enhance the robustness such that new path angles appear in the current block can still be measured with sufficient energy.

\subsection{Complexity Comparison}
The specific numbers of FLOPs and the average running time on the same CPU of different approaches under different system settings are shown in Table \ref{complexity}. Although the per-iteration complexity of the proposed DL-based approach is higher than the original SBL algorithm with the extra DNN part, its overall complexity is much lower than SBL thanks to much fewer layers (or iterations) required for convergence. Besides, the additional complexity in the multi-block approach is only marginal. Consistently, the average running time of DL-based approaches are also much shorter than SBL. Under the typical system setting, the multi-block DL-based approach takes less than 1 second while SBL takes near 28 seconds\footnote{The running time can be dramatically decreased with dedicated hardware like FPGA or on chips in practice\cite{svbi}.}. Therefore, the superiority of the proposed approach is demonstrated in both performance and complexity, making it a very promising solution for real time channel estimation in practical HAD massive MIMO systems.
\begin{table}[!htb]
\begin{tabular}{|c|c|c|c|c|}
\hline
\diagbox{$M_T, M_R, G$}{Approach} & SBL & Single-block DL & Multi-block DL\\ \hline
12, 12, 48 & 0.459, 4.996 & 0.014, 0.270 & 0.014, 0.296\\ \hline
16, 16, 48 & 1.450, 14.628 & 0.044, 0.549 & 0.044, 0.557\\ \hline
16, 16, 64 & 2.577, 27.827 & 0.078, 0.891 & 0.079, 0.913\\ \hline
\end{tabular}
\centering
\caption{Comparison of different approaches' complexity under different system settings. 100 iterations are executed in SBL. The two numbers in each entry of the table denote the TFLOPs and the average running time in the unit of seconds, respectively.}
\label{complexity}
\end{table}

\section{Conclusion}
In this paper, we have proposed a DL-based approach for wideband HAD mmWave massive channel estimation. The SBL algorithm is deeply unfolded and a tailored DNN is used to update Gaussian variance parameters in each SBL layer. The extension to the multi-block case has also been made. With effective exploitation of channel sparsity structures in various domains and the optimized or predicted measurement matrix, the proposed approaches demonstrate superiority over existing approaches in terms of both performance and complexity, and the strong generalization ability further enhances the practicality.


\end{document}